\documentclass[prd,aps,showpacs,epsf,floats,onecolumn]{revtex4}%
\usepackage{amssymb}
\usepackage{amsfonts}
\usepackage{amsmath}
\usepackage{graphicx}%
\providecommand{\U}[1]{\protect\rule{.1in}{.1in}}

\begin{document}
\title{\textbf{Continuous-time quantum search and time-dependent two-level quantum
systems}}
\author{\textbf{Carlo Cafaro}$^{1}$ and \textbf{Paul M.\ Alsing}$^{2}$}
\affiliation{$^{1}$SUNY Polytechnic Institute, 12203 Albany, New York, USA}
\affiliation{$^{2}$Air Force Research Laboratory, Information Directorate, 13441 Rome, New
York, USA}

\begin{abstract}
It was recently emphasized by Byrnes, Forster, and Tessler [Phys. Rev. Lett.
120, 060501 (2018)] that the continuous-time formulation of Grover's quantum
search algorithm can be intuitively understood in terms of Rabi oscillations
between the source and the target subspaces.

In this work, motivated by this insightful remark and starting from the
consideration of a time-independent generalized quantum search Hamiltonian as
originally introduced by Bae and Kwon [Phys. Rev. A 66, 012314 (2002)], we
present a detailed investigation concerning the physical connection between
quantum search Hamiltonians and exactly solvable time-dependent two-level
quantum systems. Specifically, we compute in an exact analytical manner the
transition probabilities from a source state to a target state in a number of
physical scenarios specified by a spin-$1/2$ particle immersed in an external
time-dependent magnetic field. In particular, we analyze both the periodic
oscillatory as well as the monotonic temporal behaviors of such transition
probabilities and, moreover, explore their analogy with characteristic
features of Grover-like and fixed-point quantum search algorithms,
respectively. Finally, we discuss from a physics standpoint the connection
between the schedule of a search algorithm, in both adiabatic and nonadiabatic
quantum mechanical evolutions, and the control fields in a time-dependent
driving Hamiltonian.

\end{abstract}

\maketitle

\bigskip\pagebreak

\section{Introduction}

The first continuous-time version of Grover's original discrete quantum search
algorithm was proposed by Farhi and Gutmann in Ref. \cite{farhi98}. In
particular, a generalization of the Farhi-Gutmann analog quantum search
algorithm was considered by Bae and Kwon in Ref. \cite{bae02}. Both quantum
search Hamiltonians in Refs. \cite{farhi98,bae02} are time-independent. The
transition from time-independence to time-dependent in the framework of analog
quantum search algorithms appears originally under the working assumption of
global adiabatic evolution in Ref. \cite{farhi00} and later, in the more
advantageous setting of local adiabaticity in Ref. \cite{roland02}. In the
local adiabatic evolution approach, the time-dependent search Hamiltonian is
the linear interpolation between two time-independent Hamiltonians.
Specifically, the system is prepared in the ground state of one of the two
Hamiltonians. Then, this Hamiltonian is adiabatically changed into the other
one whose ground state is assumed to encode the unknown solution of the search
problem. The time-dependence of the Hamiltonian is encoded in the
time-dependent interpolation function. In particular, the precise form of such
a function can be determined by imposing the local adiabaticity condition at
each instant of time during the quantum mechanical evolution of the system.
Roughly speaking, the adiabatic approximation states that a system prepared in
an instantaneous eigenstate of the Hamiltonian will stay close to this
prepared state if the Hamiltonian changes sufficiently slowly
\cite{sanders04,oh05}. Therefore, any functional form of the interpolation
schedule that satisfies the necessary degree of slowness for the adiabatic
condition to be fulfilled can be formally considered. We emphasize
that\textbf{ }there is no fundamental reason to exclude a nonlinear
interpolation of the two static Hamiltonians \cite{ali04}. Indeed, Perez and
Romanelli considered nonadiabatic quantum search Hamiltonians specified in
terms of a nonlinear interpolation of the two time-independent Hamiltonians in
Ref. \cite{romanelli07}. Depending on both the parametric and temporal form of
the two interpolation functions used to define the time-dependent Hamiltonian,
they presented two search algorithms exhibiting different features. In
particular, while both of them exhibited the typical $\mathcal{O}\left(
\sqrt{N}\right)  $ Grover-like scaling behavior in finding the target state in
an $N$-dimensional search space, one of the algorithms required a specific
time to make the measurement and find the target state (periodic oscillatory
behavior and original Grover search algorithm,
\cite{grover97,cafaro12a,cafaro12b}). The other one, instead, caused the
source state to evolve asymptotically into a quantum state that overlapped
with the target state with high probability (monotonic behavior and
fixed-point Grover search algorithm, \cite{grover05,cafaro17}).

Recently, Dolzell, Yoder, and Chuang investigated in detail the possibility
for adiabatic quantum search algorithms to exhibit the fixed-point property in
addition to quadratic quantum speedup \cite{dalzell17}. From their theoretical
analysis, supported by a number of illustrative examples, they concluded that
depending on the choice of the interpolation schedule of the algorithm, it is
possible to construct quantum search Hamiltonians with a variety of features:
Grover-like scaling and fixed-point property; Grover-like scaling but no
fixed-point property;\ fixed-point property but no Grover-like scaling. It is
especially relevant to our present work to recognize\textbf{ }that their
theoretical investigation was deeply influenced by the intuition that a
two-dimensional system whose dynamics is governed by a time-dependent
Hamiltonian is essentially equivalent to a spin-$1/2$ particle subject to an
external time-dependent magnetic field. This link between the quantum dynamics
of a spin-$1/2$ particle in an external magnetic field and quantum search
algorithms was further discussed by Byrnes, Forster, and Tessler in Ref.
\cite{byrnes18}. In this work, the authors point out that Grover's original
quantum search algorithm occurs in the setting of discrete variable quantum
computing and, in particular, the Grover operator only inverts the sign to one
state. In the framework of quantum computing with continuous variables where
the units of quantum information are called \emph{qunats} \cite{braunstein05},
changing a phase to simply a single quantum subspace of an
infinite-dimensional Hilbert space implies that the Grover operator would have
to invert the phase of an infinitely squeezed momentum state \cite{pati00}.
This task presents two drawbacks: first, it is difficult to prepare such a
state from an experimental standpoint; second, this infinitely squeezed
momentum state would have no quantum mechanical overlap with the solution
states encoded in the position eigenstates. Motivated by these considerations,
Byrnes, Forster, and Tessler present in Ref. \cite{byrnes18} a continuous-time
generalization of Grover's original quantum search algorithm where not only
the Oracle operator inverts the phase of an arbitrary number of target states
but the Grover operator also inverts the sign to an arbitrary number of
quantum states. In particular, they pointed out that Grover's search
Hamiltonian can be intuitively understood in terms of Rabi oscillations
between the source and the target subspaces.

In this paper, inspired by the analogy discussed in Ref. \cite{byrnes18}, we
seek to provide a unifying perspective on the above mentioned different types
of analog quantum search algorithms in terms of the quantum mechanics of
two-level quantum systems. In particular, beginning from the analysis of a
time-independent generalized quantum search Hamiltonian as originally
introduced in Ref. \cite{bae02}, we present a detailed investigation
concerning the physical connection between quantum search Hamiltonians and
exactly solvable time-dependent, two-level quantum systems. Specifically, we
analytically calculate the transition probabilities from a source state to a
target state in a number of physical scenarios specified by a spin-$1/2$
particle immersed in an external time-dependent magnetic field. In particular,
we analyze the observed periodic oscillatory and monotonic temporal behaviors
of such transition probabilities and investigate their analogy with
characterizing features of Grover-like and fixed-point quantum search
algorithms, respectively. Finally, we discuss the physical link between the
schedule of a search algorithm, in both adiabatic and nonadiabatic quantum
mechanical evolutions, and the control fields in a driving time-dependent Hamiltonian.

The layout of the remainder of the paper is as follows. In Section II, we
examine the transition probability from a source state to a target state for a
two-level quantum system whose time-independent Hamiltonian evolution was
originally considered by Bae and Kwon in Ref. \cite{bae02}. In Section III, we
transition from time-independent to time-dependent Hamiltonians. Specifically,
we discuss the temporal behavior of the transition probability from a source
state to a target state for a two-level quantum system immersed in a
time-dependent external magnetic field configuration that specifies the
original scenario considered by Rabi in Ref. \cite{rabi37}. The explicit
computations leading to uncover the exact temporal behavior of both quantum
mechanical probability amplitudes and transition probabilities as presented in
Sections II and III (and, in a more detailed fashion, in Appendices A and C,
respectively) allow us to strengthen the original intuitions reported in Refs.
\cite{dalzell17,byrnes18}. These results also enable a transparent comparison
between the dynamics generated by a time-independent Grover-like quantum
search Hamiltonian and a time-dependent Rabi-like two-level quantum
Hamiltonian. In Section IV, we present the basic notions of adiabatic and
nonadiabatic quantum search Hamiltonians. In the adiabatic case, we emphasize
the intuitive link between the dynamics generated by quantum search
Hamiltonians and the non-relativistic quantum mechanical dynamics of an
electron in an external time-dependent magnetic field. In the nonadiabatic
case, instead, we focus on the temporal behavior of the transition
probabilities from a source state to a target state under suitable working
assumptions on the schedules of the nonadiabatic search Hamiltonian. In
Section V, upon relaxing the working assumption of knowing \emph{a priori} the
exact time-dependent external magnetic field configuration, we study two
quantum dynamical evolution scenarios specified by suitable time-dependent
magnetic field configurations (determined \emph{a posteriori}) that produce
transition probabilities showing a temporal monotonic convergence from the
source to the target states. In the former scenario we assume that the phase
and the magnitude of the transverse field are interconnected. In the latter
scenario, while preserving the same formal expression of the \emph{complex}
transversal field, we do not assume any connection between the phase and the
intensity of the field itself. Finally, exploiting the analogies between the
dynamics generated by a time-independent Grover-like quantum search
Hamiltonian and a time-dependent Rabi-like two-level quantum Hamiltonian
presented in Sections II and III, a comparison between the dynamics of the
newly selected time-dependent Hamiltonians and that of nonadiabatic quantum
search Hamiltonians with a fixed-point property is presented. Our final
remarks appear in Section VI. Several technical details are presented in the
Appendices A, B, C, D, and E.

\section{A time-independent quantum search Hamiltonian}

Consider the time-independent generalized quantum search (GQS) Hamiltonian
$\mathcal{H}_{\text{GQS}}$ defined as \cite{bae02},%
\begin{equation}
\mathcal{H}_{\text{GQS}}\overset{\text{def}}{=}E\left[  \alpha\left\vert
w\right\rangle \left\langle w\right\vert +\beta\left\vert w\right\rangle
\left\langle s\right\vert +\gamma\left\vert s\right\rangle \left\langle
w\right\vert +\delta\left\vert s\right\rangle \left\langle s\right\vert
\right]  \text{,} \label{hamilton}%
\end{equation}
where $\alpha$, $\beta$, $\gamma$, $\delta$ are \emph{complex} expansion
coefficients. Furthermore, assume that the quantum state $\left\vert
w\right\rangle $ is the normalized target state while $\left\vert
s\right\rangle $ is the normalized initial state with \emph{real} quantum
overlap $\left\langle w|s\right\rangle =\left\langle s|w\right\rangle =x$ that
evolves in an unitary fashion according to Schr\"{o}dinger's quantum
mechanical evolution law,%
\begin{equation}
\left\vert s\right\rangle \mapsto e^{-\frac{i}{\hslash}\mathcal{H}%
_{\text{GQS}}t}\left\vert s\right\rangle \text{,}%
\end{equation}
with $\hslash\simeq0.66\times10^{-15}\left(  \text{eV}\cdot\sec\right)  $
denoting the reduced Planck constant. Our goal is to compute the time
$t^{\ast}$ such that $\mathcal{P}_{\left\vert s\right\rangle \rightarrow
\left\vert w\right\rangle }\left(  t^{\ast}\right)  =\mathcal{P}_{\max}$ where
$\mathcal{P}_{\left\vert s\right\rangle \rightarrow\left\vert w\right\rangle
}\left(  t\right)  $ is the transition probability defined as,%
\begin{equation}
\mathcal{P}_{\left\vert s\right\rangle \rightarrow\left\vert w\right\rangle
}\left(  t\right)  \overset{\text{def}}{=}\left\vert \left\langle
w|e^{-\frac{i}{\hslash}\mathcal{H}_{\text{GQS}}t}|s\right\rangle \right\vert
^{2}\text{.} \label{fidelity}%
\end{equation}
Employing the Gram-Schmidt orthonormalization technique, we can construct an
orthonormal set of state vectors starting from the set $\left\{  \left\vert
w\right\rangle \text{, }\left\vert s\right\rangle \right\}  $. We obtain,
\begin{equation}
\left\{  \left\vert w\right\rangle \text{, }\left\vert s\right\rangle
\right\}  \rightarrow\left\{  \left\vert w\right\rangle \text{, }\left\vert
s\right\rangle -\left\langle s|w\right\rangle \left\vert w\right\rangle
\right\}  \rightarrow\left\{  \frac{\left\vert w\right\rangle }{\left\Vert
\left\vert w\right\rangle \right\Vert }\text{, }\frac{\left\vert
s\right\rangle -\left\langle s|w\right\rangle \left\vert w\right\rangle
}{\left\Vert \left\vert s\right\rangle -\left\langle s|w\right\rangle
\left\vert w\right\rangle \right\Vert }\right\}  \text{.}%
\end{equation}
Let us define the quantum state vector $\left\vert r\right\rangle $ as,%
\begin{equation}
\left\vert r\right\rangle \overset{\text{def}}{=}\frac{\left\vert
s\right\rangle -\left\langle s|w\right\rangle \left\vert w\right\rangle
}{\left\Vert \left\vert s\right\rangle -\left\langle s|w\right\rangle
\left\vert w\right\rangle \right\Vert }\text{.} \label{erre2}%
\end{equation}
Recalling that $\left\langle s|w\right\rangle =x$, Eq. (\ref{erre2}) becomes%
\begin{equation}
\left\vert r\right\rangle =\frac{\left\vert s\right\rangle -\left\langle
s|w\right\rangle \left\vert w\right\rangle }{\sqrt{\left\langle
s|s\right\rangle -\left\langle s|w\right\rangle ^{2}}}=\frac{1}{\sqrt{1-x^{2}%
}}\left(  \left\vert s\right\rangle -x\left\vert w\right\rangle \right)
\text{.}%
\end{equation}
In terms of the orthonormal basis $\left\{  \left\vert w\right\rangle \text{,
}\left\vert r\right\rangle \right\}  $, the state $\left\vert s\right\rangle $
becomes%
\begin{equation}
\left\vert s\right\rangle =\left\vert s\right\rangle \left(  \left\vert
w\right\rangle \left\langle w\right\vert +\left\vert r\right\rangle
\left\langle r\right\vert \right)  =\left\langle w|s\right\rangle \left\vert
w\right\rangle +\left\langle r|s\right\rangle \left\vert r\right\rangle
\text{.} \label{chell}%
\end{equation}
Note that the quantum overlap $\left\langle r|s\right\rangle $ is given by,%
\begin{equation}
\left\langle r|s\right\rangle =\frac{1}{\sqrt{1-x^{2}}}\left(  \left\langle
s\right\vert -x\left\langle w\right\vert \right)  \left(  \left\vert
s\right\rangle \right)  =\frac{1}{\sqrt{1-x^{2}}}\left(  1-x^{2}\right)
=\sqrt{1-x^{2}}\text{.} \label{chist}%
\end{equation}
Therefore, by\textbf{ }employing Eq. (\ref{chist}), the state $\left\vert
s\right\rangle $ in Eq. (\ref{chell}) can be written as
\begin{equation}
\left\vert s\right\rangle =x\left\vert w\right\rangle +\sqrt{1-x^{2}%
}\left\vert r\right\rangle \text{.} \label{sr}%
\end{equation}
After a straightforward but tedious computation, we find that the transition
probability $P_{\left\vert s\right\rangle \rightarrow\left\vert w\right\rangle
}\left(  t\right)  $ in Eq. (\ref{fidelity}) becomes%
\begin{equation}
\mathcal{P}_{\left\vert s\right\rangle \rightarrow\left\vert w\right\rangle
}\left(  t\right)  =x^{2}\cos^{2}\left(  \sqrt{\frac{h_{12}h_{21}}{\hslash
^{2}}+\frac{\left(  h_{11}-h_{22}\right)  ^{2}}{4\hslash^{2}}}t\right)
+\frac{\left\vert \frac{1}{2}\frac{h_{11}-h_{22}}{\hslash}x+\frac{h_{12}%
}{\hslash}\sqrt{1-x^{2}}\right\vert ^{2}}{\frac{h_{12}h_{21}}{\hslash^{2}%
}+\frac{\left(  h_{11}-h_{22}\right)  ^{2}}{4\hslash^{2}}}\sin^{2}\left(
\sqrt{\frac{h_{12}h_{21}}{\hslash^{2}}+\frac{\left(  h_{11}-h_{22}\right)
^{2}}{4\hslash^{2}}}t\right)  \text{,} \label{it2}%
\end{equation}
where $h_{11}$, $h_{12}$, $h_{21}$, and $h_{22}$ are defined as,
\begin{align}
&  h_{11}\overset{\text{def}}{=}E\left[  \alpha+\left(  \beta+\gamma\right)
x+\delta x^{2}\right]  \text{, }h_{12}\overset{\text{def}}{=}E\sqrt{1-x^{2}%
}\left(  \beta+\delta x\right)  \text{,}\nonumber\\
& \nonumber\\
&  h_{21}\overset{\text{def}}{=}E\sqrt{1-x^{2}}\left(  \gamma+\delta x\right)
\text{, }h_{22}\overset{\text{def}}{=}E\delta\left(  1-x^{2}\right)  \text{.}
\label{heq}%
\end{align}
A detailed derivation of Eq. (\ref{it2}) appears in Appendix A. From Eq.
(\ref{it2}), it follows that the maximum $\mathcal{P}_{\max}=\mathcal{P}%
_{\left\vert s\right\rangle \rightarrow\left\vert w\right\rangle }\left(
t^{\ast}\right)  $ of $\mathcal{P}_{\left\vert s\right\rangle \rightarrow
\left\vert w\right\rangle }\left(  t\right)  $ occurs at the instant $t^{\ast
}$,%
\begin{equation}
t^{\ast}\overset{\text{def}}{=}\frac{\pi\hslash}{\sqrt{\left(  h_{11}%
-h_{22}\right)  ^{2}+4h_{12}h_{21}}}\text{,} \label{tmax}%
\end{equation}
and, in addition, is equal to%
\begin{equation}
\mathcal{P}_{\max}=\frac{\left\vert \frac{1}{2}\frac{h_{11}-h_{22}}{\hslash
}x+\frac{h_{12}}{\hslash}\sqrt{1-x^{2}}\right\vert ^{2}}{\frac{h_{12}h_{21}%
}{\hslash^{2}}+\frac{\left(  h_{11}-h_{22}\right)  ^{2}}{4\hslash^{2}}%
}\text{.} \label{maxp}%
\end{equation}
Finally, using\ Eq.(\ref{heq}) and observing that $\alpha$ and $\delta$ must
be \emph{real} coefficients while $\beta=\gamma^{\ast}$, $t^{\ast}$ in Eq.
(\ref{tmax}) and $\mathcal{P}_{\max}$ in Eq. (\ref{maxp}) become%
\begin{equation}
t^{\ast}\left(  \alpha\text{, }\beta\text{, }\delta\text{, }x\right)
\overset{\text{def}}{=}\frac{2}{\sqrt{4\left[  \alpha\delta+\operatorname{Re}%
^{2}\left(  \beta\right)  -\left\vert \beta\right\vert ^{2}\right]
x^{2}+4\operatorname{Re}\left(  \beta\right)  \left(  \alpha+\delta\right)
x+\left[  \left(  \alpha-\delta\right)  ^{2}+4\left\vert \beta\right\vert
^{2}\right]  }}\frac{\pi\hslash}{2E}\text{,}%
\end{equation}
and,%
\begin{equation}
\mathcal{P}_{\max}\left(  \alpha\text{, }\beta\text{, }\delta\text{,
}x\right)  =\frac{4\left[  \left\vert \beta\right\vert ^{2}-\operatorname{Re}%
^{2}\left(  \beta\right)  \right]  x^{4}+\left[  \left(  \alpha+\delta\right)
^{2}-8\left(  \left\vert \beta\right\vert ^{2}-\operatorname{Re}^{2}\left(
\beta\right)  \right)  \right]  x^{2}+4\operatorname{Re}\left(  \beta\right)
\left(  \alpha+\delta\right)  x+4\left\vert \beta\right\vert ^{2}}{4\left[
\alpha\delta+\operatorname{Re}^{2}\left(  \beta\right)  -\left\vert
\beta\right\vert ^{2}\right]  x^{2}+4\operatorname{Re}\left(  \beta\right)
\left(  \alpha+\delta\right)  x+\left[  \left(  \alpha-\delta\right)
^{2}+4\left\vert \beta\right\vert ^{2}\right]  }\text{,}%
\end{equation}
respectively. For the sake of completeness, we point out that in the limiting
case of $\alpha=\delta=1$ and $\gamma=\beta^{\ast}=0$, we recover the original
Farhi-Gutmann (FG) results,%
\begin{equation}
t_{\text{FG}}^{\ast}=\frac{\pi\hslash}{2E}\cdot\frac{1}{x}\text{, and
}\mathcal{P}_{\max}^{\left(  \text{FG}\right)  }=1\text{.} \label{GF}%
\end{equation}
Moreover, we also recover the transition probability $\mathcal{P}_{\left\vert
s\right\rangle \rightarrow\left\vert w\right\rangle }^{\left(  \text{GF}%
\right)  }\left(  t\right)  $,
\begin{equation}
\mathcal{P}_{\left\vert s\right\rangle \rightarrow\left\vert w\right\rangle
}^{\left(  \text{FG}\right)  }\left(  t\right)  \overset{\text{def}}{=}%
x^{2}\cos^{2}\left(  \frac{Ex}{\hslash}t\right)  +\sin^{2}\left(  \frac
{Ex}{\hslash}t\right)  \text{.}%
\end{equation}
It is insightful to point out that the minimum search time $t^{\ast}$ in Eq.
(\ref{tmax}) needed to achieve the maximal success probability $\mathcal{P}%
_{\max}$ in Eq. (\ref{maxp}) is inversely proportional to the energy level
separation $\Delta\lambda\overset{\text{def}}{=}\lambda_{+}-\lambda_{-}$ where
$\lambda_{\pm}$ denote the two \emph{real} eigenvalues of the Hermitian matrix
$\left[  \mathcal{H}_{\text{GQS}}\right]  _{\left\{  \left\vert w\right\rangle
\text{, }\left\vert r\right\rangle \right\}  }$,%
\begin{equation}
\lambda_{\pm}\overset{\text{def}}{=}\frac{1}{2}\left[  \left(  h_{11}%
+h_{22}\right)  \pm\sqrt{\left(  h_{11}-h_{22}\right)  ^{2}+4h_{12}h_{21}%
}\right]  \text{.}%
\end{equation}
Therefore, it appears reasonable from our analysis that the finiteness of the
minimum search time $t^{\ast}$ can be explained in terms of the familiar
phenomenon of avoided level crossing in quantum perturbation theory
\cite{landau}. Such a phenomenon is caused by the effect of an external
perturbation on a two-level quantum system. Specifically, the two energy
levels move apart in opposite direction after applying the perturbation. The
higher level increases while the lower level decreases. Furthermore, the
energy splitting is more prominent when the unperturbed Hamiltonian is
degenerate. This link between analog quantum search algorithms and
perturbations in quantum theory will be further explored in the next sections.

\section{Periodic oscillatory behavior and the Rabi time-dependent
Hamiltonian}

In this section, we begin by computing the transition probability from a
source state $\left\vert s\right\rangle $ to a target state $\left\vert
w\right\rangle $ for a two-level quantum system immersed in a time-dependent
external magnetic field configuration that characterizes the original scenario
considered by Rabi in Ref. \cite{rabi37}. For a brief review on the concept of
interaction representation in quantum mechanics in the context of
time-dependent perturbations, we refer to Appendix B.

\subsection{Transition probability: General setting}

The general expression of the time-dependent Hamiltonian of the system that we
are investigating is given by,%
\begin{equation}
\mathcal{H}^{\left(  \text{full}\right)  }\left(  t\right)  \overset
{\text{def}}{=}\mathcal{H}_{0}^{\left(  \text{free}\right)  }+\mathcal{V}%
^{\left(  \text{interaction}\right)  }\left(  t\right)  \text{.} \label{rabi1}%
\end{equation}
In particular, to study the original scenario considered by Rabi in Ref.
\cite{rabi37}, assume that the time-independent free Hamiltonian
$\mathcal{H}_{0}^{\left(  \text{free}\right)  }$ and the time-dependent
interaction potential $\mathcal{V}^{\left(  \text{interaction}\right)
}\left(  t\right)  $ are defined as \cite{sakurai},%
\begin{equation}
\mathcal{H}_{0}^{\left(  \text{free}\right)  }\overset{\text{def}}{=}%
E_{1}\left\vert E_{1}\right\rangle \left\langle E_{1}\right\vert
+E_{2}\left\vert E_{2}\right\rangle \left\langle E_{2}\right\vert \text{, and
}\mathcal{V}^{\left(  \text{interaction}\right)  }\left(  t\right)
\overset{\text{def}}{=}\Gamma e^{i\omega t}\left\vert E_{1}\right\rangle
\left\langle E_{2}\right\vert +\Gamma e^{-i\omega t}\left\vert E_{2}%
\right\rangle \left\langle E_{1}\right\vert \text{,} \label{rabi2}%
\end{equation}
respectively. The two parameters $\omega$ and $\Gamma$ are positive and
\emph{real}. Furthermore, we assume that $E_{2}>E_{1}$ and $\left\langle
E_{i}|E_{j}\right\rangle =\delta_{ij}$ for any $1\leq i$, $j\leq2$. Using Eqs.
(\ref{odesystem}) and (\ref{rabi2}), we obtain%
\begin{equation}
i\hslash\left(
\begin{array}
[c]{c}%
\dot{c}_{1}\\
\dot{c}_{2}%
\end{array}
\right)  =\left(
\begin{array}
[c]{cc}%
\mathcal{V}_{11}\left(  t\right)  & \mathcal{V}_{12}\left(  t\right)
e^{i\omega_{12}t}\\
\mathcal{V}_{21}\left(  t\right)  e^{i\omega_{21}t} & \mathcal{V}_{22}\left(
t\right)
\end{array}
\right)  \left(
\begin{array}
[c]{c}%
c_{1}\\
c_{2}%
\end{array}
\right)  \text{,}%
\end{equation}
that is,%
\begin{equation}
i\hslash\left(
\begin{array}
[c]{c}%
\dot{c}_{1}\\
\dot{c}_{2}%
\end{array}
\right)  =\left(
\begin{array}
[c]{cc}%
0 & \Gamma e^{i\left(  \omega-\omega_{21}\right)  t}\\
\Gamma e^{-i\left(  \omega-\omega_{21}\right)  t} & 0
\end{array}
\right)  \left(
\begin{array}
[c]{c}%
c_{1}\\
c_{2}%
\end{array}
\right)  \text{,} \label{system}%
\end{equation}
where $\dot{c}_{i}\overset{\text{def}}{=}dc_{i}/dt$ for any $1\leq i$,
$j\leq2$ and $\omega_{21}\overset{\text{def}}{=}\left(  E_{2}-E_{1}\right)
/\hslash$ denotes the characteristic frequency of the two-level system. The
matrix equality in Eq. (\ref{system})\ yields a system of two coupled first
order ordinary differential equations,%
\begin{equation}
i\hslash\dot{c}_{1}=\Gamma e^{i\left(  \omega-\omega_{21}\right)  t}%
c_{2}\text{, and }i\hslash\dot{c}_{2}=\Gamma e^{-i\left(  \omega-\omega
_{21}\right)  t}c_{1}\text{.} \label{system1}%
\end{equation}

\begin{figure}[t]
\centering
\includegraphics[width=0.35\textwidth] {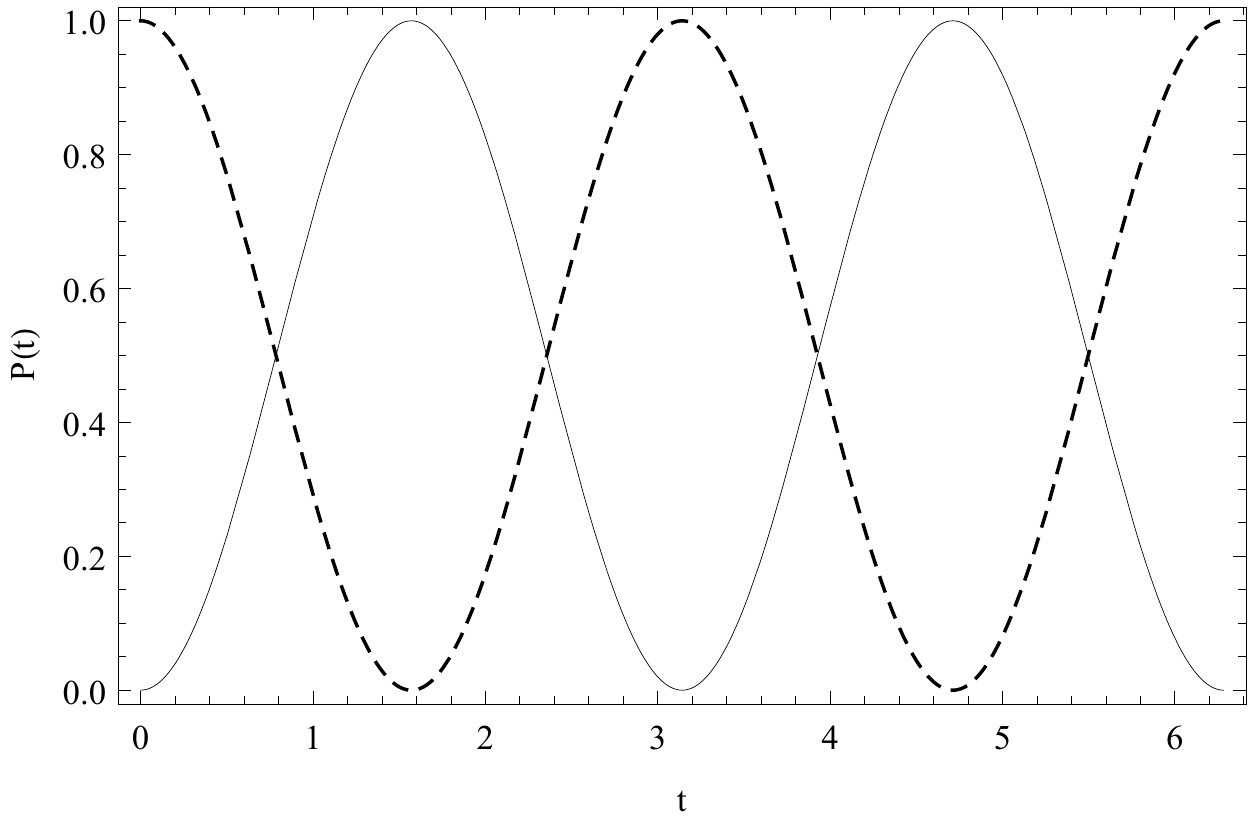}\caption{Oscillatory and periodic
temporal behavior of the transition probabilities $\mathcal{P}_{\left\vert
s\right\rangle \rightarrow\left\vert E_{2}\right\rangle }\left(  t\right)  $
(solid line) and $\mathcal{P}_{\left\vert s\right\rangle \rightarrow\left\vert
E_{1}\right\rangle }\left(  t\right)  $ (dashed line) at the original Rabi
resonance where $\omega=\omega_{21}$ with $x=0$, $\Gamma=1$, and $\hslash=1$.}%
\label{fig1}%
\end{figure}At this point, we recall that the transition probabilities
$\mathcal{P}_{\left\vert s\right\rangle \rightarrow\left\vert E_{1}%
\right\rangle }\left(  t\right)  $ and $\mathcal{P}_{\left\vert s\right\rangle
\rightarrow\left\vert E_{2}\right\rangle }\left(  t\right)  $ are defined as
$\mathcal{P}_{\left\vert s\right\rangle \rightarrow\left\vert E_{1}%
\right\rangle }\left(  t\right)  \overset{\text{def}}{=}\left\vert
c_{1}\left(  t\right)  \right\vert ^{2}$ and $\mathcal{P}_{\left\vert
s\right\rangle \rightarrow\left\vert E_{2}\right\rangle }\left(  t\right)
\overset{\text{def}}{=}\left\vert c_{2}\left(  t\right)  \right\vert ^{2}$,
respectively. After some algebra (for more details, we refer to Appendix C),
we find that the transition probabilities $\mathcal{P}_{\left\vert
s\right\rangle \rightarrow\left\vert E_{1}\right\rangle }\left(  t\right)  $
and $\mathcal{P}_{\left\vert s\right\rangle \rightarrow\left\vert
E_{2}\right\rangle }\left(  t\right)  $ are given by%
\begin{equation}
\mathcal{P}_{\left\vert s\right\rangle \rightarrow\left\vert E_{1}%
\right\rangle }\left(  t\right)  =\left(  1-x^{2}\right)  \cos^{2}\left(
\sqrt{\frac{\Gamma^{2}}{\hslash^{2}}+\frac{\left(  \omega-\omega_{21}\right)
^{2}}{4}}t\right)  +\left\{  1-\left[  \frac{\frac{\left(  \omega-\omega
_{21}\right)  }{2}x-\frac{\Gamma}{\hslash}\sqrt{1-x^{2}}}{\sqrt{\frac
{\Gamma^{2}}{\hslash^{2}}+\frac{\left(  \omega-\omega_{21}\right)  ^{2}}{4}}%
}\right]  ^{2}\right\}  \sin^{2}\left(  \sqrt{\frac{\Gamma^{2}}{\hslash^{2}%
}+\frac{\left(  \omega-\omega_{21}\right)  ^{2}}{4}}t\right)  \text{,}
\label{P1}%
\end{equation}
and,%
\begin{equation}
\mathcal{P}_{\left\vert s\right\rangle \rightarrow\left\vert E_{2}%
\right\rangle }\left(  t\right)  =x^{2}\cos^{2}\left(  \sqrt{\frac{\Gamma^{2}%
}{\hslash^{2}}+\frac{\left(  \omega-\omega_{21}\right)  ^{2}}{4}}t\right)
+\left[  \frac{\frac{\left(  \omega-\omega_{21}\right)  }{2}x-\frac{\Gamma
}{\hslash}\sqrt{1-x^{2}}}{\sqrt{\frac{\Gamma^{2}}{\hslash^{2}}+\frac{\left(
\omega-\omega_{21}\right)  ^{2}}{4}}}\right]  ^{2}\sin^{2}\left(  \sqrt
{\frac{\Gamma^{2}}{\hslash^{2}}+\frac{\left(  \omega-\omega_{21}\right)  ^{2}%
}{4}}t\right)  \text{,} \label{P2}%
\end{equation}
respectively. In Fig. 1, we report the oscillatory and periodic temporal
behavior of the transition probabilities $\mathcal{P}_{\left\vert
s\right\rangle \rightarrow\left\vert E_{2}\right\rangle }\left(  t\right)  $
and $\mathcal{P}_{\left\vert s\right\rangle \rightarrow\left\vert
E_{1}\right\rangle }\left(  t\right)  $ at the original Rabi resonance where
$\omega=\omega_{21}$ with $x=0$, $\Gamma=1$, and $\hslash=1$. For the sake of
completeness, observe that%
\begin{equation}
_{\text{I}}\left\langle s\left(  t\right)  |s\left(  t\right)  \right\rangle
_{\text{I}}=\mathcal{P}_{\left\vert s\right\rangle \rightarrow\left\vert
E_{1}\right\rangle }\left(  t\right)  +\mathcal{P}_{\left\vert s\right\rangle
\rightarrow\left\vert E_{2}\right\rangle }\left(  t\right)  =1\text{,}%
\end{equation}
where the quantum state $\left\vert s\left(  t\right)  \right\rangle
_{\text{I}}$ is such that,%
\begin{equation}
\left\vert s\left(  t\right)  \right\rangle _{\text{I}}\overset{\text{def}}%
{=}\left\langle E_{1}|s\left(  t\right)  \right\rangle _{\text{I}}\left\vert
E_{1}\right\rangle +\left\langle E_{2}|s\left(  t\right)  \right\rangle
_{\text{I}}\left\vert E_{2}\right\rangle \text{,}%
\end{equation}
with $\left\vert s\left(  0\right)  \right\rangle _{\text{I}}\overset
{\text{def}}{=}\sqrt{1-x^{2}}\left\vert E_{1}\right\rangle +x\left\vert
E_{2}\right\rangle $.

\subsection{Time-dependent magnetic field in the Rabi Hamiltonian}

At this point, in an effort to emphasize several physics insights between
analog quantum search algorithms and time-dependent Hamiltonians, we recall
that the most general time-dependent Hermitian Hamiltonian operator
$\mathcal{H}\left(  t\right)  $ with a $\left(  2\times2\right)  $-matrix
representation can be recast as,%
\begin{equation}
\mathcal{H}\left(  t\right)  \overset{\text{def}}{=}a\left(  t\right)
\mathcal{I}+\vec{b}\left(  t\right)  \cdot\vec{\sigma}\text{,}
\label{canonicalbs1}%
\end{equation}
where $\vec{\sigma}$ denotes the vector of Pauli matrices, $\mathcal{I}$ is
the identity operator, $a\left(  t\right)  $ is a time-dependent function, and
$\vec{b}\left(  t\right)  $ is the vector given by%
\begin{equation}
\vec{b}\left(  t\right)  \overset{\text{def}}{=}\left\Vert \vec{b}\left(
t\right)  \right\Vert \hat{b}\left(  t\right)  =\sqrt{\left[  b_{x}\left(
t\right)  \right]  ^{2}+\left[  b_{y}\left(  t\right)  \right]  ^{2}+\left[
b_{z}\left(  t\right)  \right]  ^{2}}\hat{b}\left(  t\right)  \text{.}%
\end{equation}
Recalling that the Pauli vector operator $\vec{\sigma}$ is defined as
\cite{cafaro10,cafaro14},%
\begin{equation}
\vec{\sigma}\overset{\text{def}}{=}\left[  \sigma_{x}\text{, }\sigma
_{y}\text{, }\sigma_{z}\right]  =\left[  \left(
\begin{array}
[c]{cc}%
0 & 1\\
1 & 0
\end{array}
\right)  \text{,}\left(
\begin{array}
[c]{cc}%
0 & -i\\
i & 0
\end{array}
\right)  \text{, }\left(
\begin{array}
[c]{cc}%
1 & 0\\
0 & -1
\end{array}
\right)  \right]  \text{,}%
\end{equation}
we note that the matrix representation $\left[  \mathcal{H}\left(  t\right)
\right]  _{\mathcal{B}_{\text{canonical}}}$ of the Hamiltonian $\mathcal{H}%
\left(  t\right)  $ in Eq. (\ref{canonicalbs1}) with respect to the canonical
basis $\mathcal{B}_{\text{canonical}}$ is given by,%
\begin{equation}
\left[  \mathcal{H}\left(  t\right)  \right]  _{\mathcal{B}_{\text{canonical}%
}}=\left(
\begin{array}
[c]{cc}%
a+b_{z} & b_{x}-ib_{y}\\
b_{x}+ib_{y} & a-b_{z}%
\end{array}
\right)  \text{.} \label{canonicalbs}%
\end{equation}
For the sake of notational simplicity, we have omitted to make explicit the
time-dependence of the scalar function $a=a\left(  t\right)  $ and the vector
$\vec{b}=\vec{b}\left(  t\right)  $ in Eq. (\ref{canonicalbs1}). The Rabi-like
Hamiltonian $\mathcal{H}^{\left(  \text{full}\right)  }\left(  t\right)  $ in
Eq. (\ref{rabi1}) can be rewritten as,%
\begin{equation}
\mathcal{H}^{\left(  \text{full}\right)  }\left(  t\right)  =-\vec{\mu}%
\cdot\vec{B}\text{,} \label{physicsH}%
\end{equation}
\ where the vector $\vec{B}$ denotes the external time-dependent magnetic
field with constant magnitude,%
\begin{equation}
\vec{B}\left(  t\right)  \overset{\text{def}}{=}\left(  B_{1}\cos\left(
\omega t\right)  \text{, }B_{1}\sin\left(  \omega t\right)  \text{, }%
B_{0}\right)  \text{,}%
\end{equation}
with $B_{1}$, $B_{0}$, and the angular frequency $\omega$ belonging to $%
\mathbb{R}
_{+}\backslash\left\{  0\right\}  $. The vector operator $\vec{\mu}$ is the
magnetic moment of the electron,%
\begin{equation}
\vec{\mu}\overset{\text{def}}{=}\frac{e\hslash}{2mc}\vec{\sigma}\text{,}%
\end{equation}
with $e$ being the electric charge of the electron, $m$ is its mass, and $c$
denotes the speed of light. Finally, the quantity $e\hslash/(2mc)\simeq
5.8\times10^{-9}\left(  \text{eV}/\text{gauss}\right)  $ denotes the Bohr magneton.

\begin{table}[t]
\centering
\begin{tabular}
[c]{c|c|c|c|c|c}\hline\hline
Hamiltonians & Energy gap & Interaction strength & Characteristic frequency &
Angular frequency & Search time\\\hline
Rabi & $\frac{\left\vert e\right\vert \hslash}{mc}B_{0}$ & $\frac{\left\vert
e\right\vert \hslash}{2mc}B_{1}$ & $\frac{\left\vert e\right\vert B_{0}}{mc}$
& $\omega$ & $\frac{\pi mc}{\left\vert e\right\vert B_{1}}$\\
Farhi-Gutmann & $2Ex^{2}$ & $Ex\sqrt{1-x^{2}}$ & $\frac{2Ex^{2}}{\hslash}$ &
$0$ & $\frac{\pi\hslash}{2Ex}$\\\hline
\end{tabular}
\caption{Schematic correspondence between selected features, including the
characteristic frequency of the system and the angular frequency of the
magnetic field vector, that characterize the quantum dynamics arising from the
Rabi and the Farhi-Gutmann Hamiltonians.}%
\end{table}We point out that $\mathcal{H}^{\left(  \text{full}\right)
}\left(  t\right)  $ in Eq. (\ref{physicsH}) is not only time-dependent but,
in general, the Hamiltonian at two different instances does not commute,%
\begin{equation}
\left[  \mathcal{H}^{\left(  \text{full}\right)  }\left(  t^{\prime}\right)
\text{, }\mathcal{H}^{\left(  \text{full}\right)  }\left(  t^{\prime\prime
}\right)  \right]  \neq0\text{,} \label{noncommutative}%
\end{equation}
where $t^{\prime}\neq t^{\prime\prime}$. For instance, taking $t^{\prime}=0$
and $t^{\prime\prime}=\pi/\left(  2\omega\right)  $ and recalling the
commutation relations between the Pauli matrices \cite{sakurai,picasso},%
\begin{equation}
\left[  \sigma_{i}\text{, }\sigma_{j}\right]  \overset{\text{def}}{=}%
\sigma_{i}\sigma_{j}-\sigma_{j}\sigma_{i}=2i\varepsilon_{ijk}\sigma
_{k}\text{,}%
\end{equation}
we obtain,%
\begin{equation}
\left[  \mathcal{H}^{\left(  \text{full}\right)  }\left(  0\right)  \text{,
}\mathcal{H}^{\left(  \text{full}\right)  }\left(  \frac{\pi}{2\omega}\right)
\right]  =\left(  \frac{e\hslash}{2mc}\right)  ^{2}2i\left[  B_{1}^{2}%
\sigma_{z}-B_{0}B_{1}\left(  \sigma_{x}+\sigma_{y}\right)  \right]
\neq0\text{.}%
\end{equation}
Fundamentally, Eq. (\ref{noncommutative}) is a consequence of the fact that in
the original Rabi scenario, although the amplitude of the magnetic field is
stationary, the direction of this field does change in time.

\subsection{Comparison between the Rabi and the Farhi-Gutmann Hamiltonians}

Having introduced the magnetic field $\vec{B}$ and the magnetic moment
$\vec{\mu}$, the connection between $\mathcal{H}^{\left(  \text{full}\right)
}\left(  t\right)  $ in Eq. (\ref{rabi1}) and $\mathcal{H}^{\left(
\text{full}\right)  }\left(  t\right)  $ in Eq. (\ref{physicsH}) becomes clear
once one considers the following correspondences,%
\begin{equation}
\omega=\omega\text{, }E_{2}=-E_{1}\overset{\text{def}}{=}-\frac{e\hslash}%
{2mc}B_{0}\text{, and }\Gamma\overset{\text{def}}{=}-\frac{e\hslash}{2mc}%
B_{1}\text{.} \label{correspondence}%
\end{equation}
A schematic correspondence between selected features that characterize the
quantum dynamics arising from the Rabi and the Farhi-Gutmann Hamiltonians
appear in Table I. In particular, equating the energy gaps and the interaction
strengths in the two frameworks, it follows that%
\begin{equation}
x=x\left(  B_{0}\text{, }B_{1}\right)  =\frac{B_{0}}{B_{1}}\frac{1}%
{\sqrt{1+\left(  \frac{B_{0}}{B_{1}}\right)  ^{2}}}\text{.} \label{chichi2}%
\end{equation}
A plot of Eq. (\ref{chichi2}) appears in Fig. $2$. Having pointed out these
correspondences, we also remark that $\mathcal{H}^{\left(  \text{full}\right)
}\left(  t\right)  $ in\ Eq. (\ref{physicsH}) can be written as%
\begin{equation}
\mathcal{H}^{\left(  \text{full}\right)  }\left(  t\right)  =\frac{1}{2}%
\Omega_{\mathcal{H}}\left(  t\right)  \hat{n}\left(  t\right)  \cdot
\vec{\sigma}\text{,} \label{fu1}%
\end{equation}
where $\Omega_{\mathcal{H}}\left(  t\right)  $ and $\hat{n}\left(  t\right)  $
are given by,
\begin{equation}
\Omega_{\mathcal{H}}\left(  t\right)  \overset{\text{def}}{=}-\frac{e\hslash
}{mc}\sqrt{\left(  B_{0}\right)  ^{2}+\left(  B_{1}\right)  ^{2}}\text{,}
\label{omegarabi1}%
\end{equation}
and,%
\begin{equation}
\hat{n}\left(  t\right)  \overset{\text{def}}{=}\frac{1}{\sqrt{\left(
B_{0}\right)  ^{2}+\left(  B_{1}\right)  ^{2}}}\left(  B_{1}\cos\left(  \omega
t\right)  \text{, }B_{1}\sin\left(  \omega t\right)  \text{, }B_{0}\right)
\text{,} \label{fu2}%
\end{equation}
respectively. Note that $\Omega_{\mathcal{H}}\left(  t\right)  $ in Eq.
(\ref{omegarabi1}) does not depend on time while the unit vector $\hat
{n}\left(  t\right)  $ becomes time-independent only if $\omega$ is set equal
to zero. In the remainder of this section, in order to facilitate discussion
of the formal comparison between Grover-like time-independent quantum search
Hamiltonians and Rabi-like time-dependent two-levels Hamiltonians, we focus
without any loss of generality on the original Farhi-Gutmann Hamiltonian. In
particular, before comparing the transitions probabilities in Eqs. (\ref{it2})
and (\ref{P2}), inspired by the Hamiltonian decomposition as presented in Eq.
(\ref{rabi1}), we note that the original Farhi-Gutmann Hamiltonian
$\mathcal{H}_{\text{FG}}^{\left(  \text{full}\right)  }$ can be formally
written as%
\begin{equation}
\mathcal{H}_{\text{FG}}^{\left(  \text{full}\right)  }\overset{\text{def}}%
{=}E\left\vert w\right\rangle \left\langle w\right\vert +E\left\vert
s\right\rangle \left\langle s\right\vert =\mathcal{H}_{\text{FG}}^{\left(
\text{free}\right)  }+\mathcal{H}_{\text{FG}}^{\left(  \text{interaction}%
\right)  }\text{,} \label{GF decomposition}%
\end{equation}
where $\left\vert s\right\rangle \overset{\text{def}}{=}\sqrt{1-x^{2}%
}\left\vert r\right\rangle +x\left\vert w\right\rangle $. In addition, the
states $\left\vert r\right\rangle $ and $\left\vert w\right\rangle $ replace
the states $\left\vert E_{1}\right\rangle $ and $\left\vert E_{2}\right\rangle
$, respectively. The Hamiltonians $\mathcal{H}_{\text{FG}}^{\left(
\text{free}\right)  }$ and $\mathcal{H}_{\text{FG}}^{\left(
\text{interaction}\right)  }$ in Eq. (\ref{GF decomposition}) are given by,%
\begin{equation}
\mathcal{H}_{\text{FG}}^{\left(  \text{free}\right)  }\overset{\text{def}}%
{=}E\left(  1-x^{2}\right)  \left\vert r\right\rangle \left\langle
r\right\vert +E\left(  1+x^{2}\right)  \left\vert w\right\rangle \left\langle
w\right\vert \label{GF1}%
\end{equation}
and,%
\begin{equation}
\mathcal{H}_{\text{FG}}^{\left(  \text{interaction}\right)  }\overset
{\text{def}}{=}Ex\sqrt{1-x^{2}}\left\vert r\right\rangle \left\langle
w\right\vert +Ex\sqrt{1-x^{2}}\left\vert w\right\rangle \left\langle
r\right\vert \text{,} \label{GF2}%
\end{equation}
respectively. From Eqs. (\ref{rabi1}), (\ref{GF1}), and (\ref{GF2}), we note
that%
\begin{equation}
E_{1}\overset{\text{def}}{=}E\left(  1-x^{2}\right)  \text{, }E_{2}%
\overset{\text{def}}{=}E\left(  1+x^{2}\right)  >E_{1}\text{, }\Gamma
\overset{\text{def}}{=}Ex\sqrt{1-x^{2}}\text{, and }\omega\overset{\text{def}%
}{=}0\text{.} \label{chichi}%
\end{equation}
From Eq. (\ref{chichi}), the energy gap $\Delta E\overset{\text{def}}{=}%
E_{2}-E_{1}=2Ex^{2}$ and, furthermore, the characteristic frequency
$\omega_{21}$ of the two-level system becomes%
\begin{equation}
\omega_{21}\overset{\text{def}}{=}\frac{\Delta E}{\hslash}=\frac{2Ex^{2}%
}{\hslash}\text{.} \label{chichi1}%
\end{equation}
In particular, using Eqs. (\ref{P2}) and (\ref{correspondence}) it can be
shown that at resonance, that is when $\omega=\omega_{21}$, the analogue of
Eq. (\ref{GF}) obtained in the original Farhi-Gutmann framework becomes%
\begin{equation}
t_{\text{Rabi}}^{\ast}=\frac{\pi\hslash}{2}\cdot\frac{1}{\Gamma}=\frac
{\pi\hslash}{2}\cdot\left(  \frac{2mc}{\left\vert e\right\vert \hslash
}\right)  \cdot\frac{1}{B_{1}}\text{, and }\mathcal{P}_{\max}^{\left(
\text{Rabi}\right)  }=1\text{.}%
\end{equation}
Having pointed out these formal correspondences between the two quantum
dynamical scenarios in Eqs. (\ref{chichi}) and (\ref{chichi1}), we conclude by
emphasizing the formally identical periodic oscillatory temporal behavior in
the transition probabilities obtained in Eqs. (\ref{it2}) and (\ref{P2}).

\begin{figure}[t]
\centering
\includegraphics[width=0.35\textwidth] {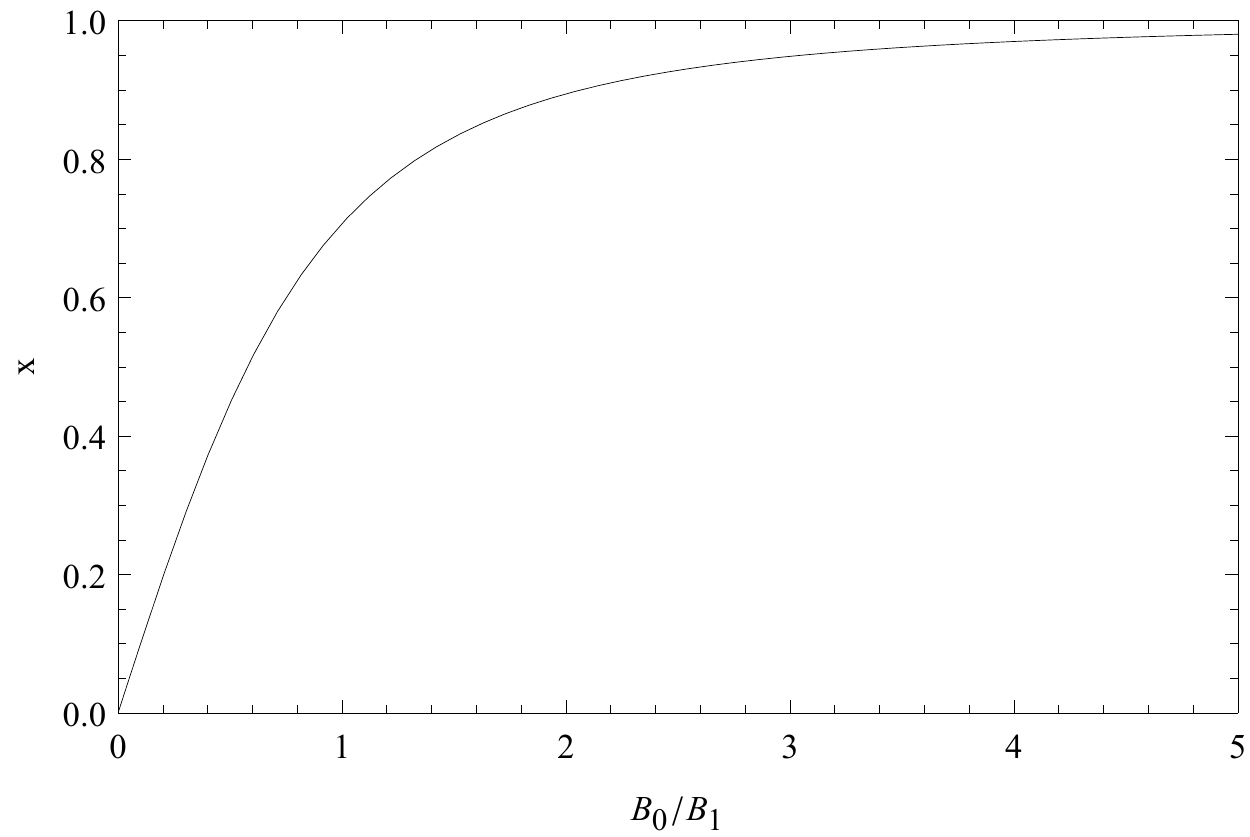}\caption{Dependence of the
quantum mechanical overlap $x$ in the Grover-like generalized quantum search
Hamiltonian setting viewed in terms of the ratio of the magnetic field
intensities $B_{0}$ and $B_{1}$ in the original Rabi-like time-dependent
Hamiltonian framework.}%
\label{fig2}%
\end{figure}

Despite these formal analogies between the Grover-like and the Rabi-like
scenarios, the spin-$1/2$ particle can be regarded as immersed in a stationary
magnetic field in the former case while it is subject to a time-dependent
magnetic field configuration in the latter one. For this reason, we observe
distinctive features in the two scenarios, both in the mathematical methods
employed (mathematical physics) and in the type of perturbation of the quantum
system (quantum physics). From a mathematical physics standpoint, instead of
using familiar matrix algebra methods and finding an explicit expression for
the unitary temporal evolution operator as we did in the first scenario, we
were compelled to employ the integration of ODEs approach to determine the
probability amplitudes in the second quantum dynamical setting specified by a
non-stationary magnetic field configuration (with constant amplitude but
time-dependent direction) assumed to be known\textbf{ }\emph{a priori}%
\textbf{. }Unfortunately, the possibility of solving in an exact analytical
fashion these systems of coupled linear ODEs with time-dependent coefficients
is rather unusual. From a quantum physics standpoint, when transitioning from
the first to the second scenario, we go from a time-independent to a
time-dependent perturbation. In the latter scenario, recalling the
considerations on the link between anticrossing and minimum search time
pointed out at the end of Section I, one can think of changing the
perturbation by suitably varying the magnetic field amplitude and/or its
frequency in order to minimize the search time. Interestingly, the passage
through resonance, either by \emph{adiabatically} varying the magnetic field
or the frequency, occurs in magnetic resonance phenomena. Thus, motivated by
the remarkable insights presented in Ref. \cite{byrnes18}, we proposed in
Sections II\ and III a link between the physics of two-level quantum systems
and analog quantum search algorithms. This link, emphasizing the role played
by perturbations, leads naturally to the consideration of the concept of
adiabaticity in\ Section IV.

The\textbf{ }next section serves as a bridge between Section III and Section
V. In the latter section, we shall investigate the possibility of finding
suitable magnetic field configurations yielding transition probabilities with
a monotonic temporal behavior that one would observe in a quantum dynamical
evolution under a fixed-point quantum search Hamiltonian. In Section IV, we
shall revisit the concepts of adiabatic and nonadiabatic quantum search algorithms.

\section{Adiabatic and nonadiabatic quantum search}

The intuitive analogy between the quantum evolution of a spin-$1/2$ particle
in an external magnetic field and the dynamics arising from a time-dependent
Hamiltonian of a two-level quantum system was originally exploited in the
context of fixed-point adiabatic quantum search by Dalzell, Yoder, and Chuang
in Ref. \cite{dalzell17}. For the sake of transparency, we point out that no
explicit temporal behavior of transition probabilities was considered by these
authors. On the contrary, such a temporal dependence was investigated in two
dynamical scenarios proposed by Perez and Romanelli in the framework of
nonadiabatic quantum search in Ref. \cite{romanelli07}. It is worth noting
however, no physical connection between the quantum search Hamiltonians and
the driven time-dependent two-level quantum systems was addressed by these
latter authors.

In what follows, motivated by our physical insights presented in Sections II
and III, we revisit the basic concepts of adiabatic and nonadiabatic quantum
search algorithms. The review of both adiabatic and nonadiabatic quantum
search algorithms together with the material that will be introduced in
Section V will help us establish a physical link between the schedule of a
search algorithm and the control fields that specify a time-dependent driving Hamiltonian.

\subsection{The adiabatic framework}

The quantum mechanical evolution in the case of adiabatic quantum search can
be described in terms of a time-dependent Hamiltonian $\mathcal{H}%
_{\text{adiabatic}}\left(  t\right)  $ specified by means of a linear
interpolation of two time-independent Hamiltonians \cite{farhi00,roland02},%
\begin{equation}
\mathcal{H}_{\text{adiabatic}}\left(  t\right)  \overset{\text{def}}{=}\left[
1-s\left(  t\right)  \right]  \mathcal{H}_{0}+s\left(  t\right)
\mathcal{H}_{1}\text{,} \label{hadiabatic}%
\end{equation}
where $\mathcal{H}_{0}\overset{\text{def}}{=}\mathcal{I-}\left\vert
s\right\rangle \left\langle s\right\vert $, $\mathcal{H}_{1}\overset
{\text{def}}{=}\mathcal{I-}\left\vert w\right\rangle \left\langle w\right\vert
$, and $\left\langle w|s\right\rangle \overset{\text{def}}{=}x\in\left(
0\text{, }1\right)  $. For the sake of completeness, we remark that there is
no fundamental reason why one could not consider a nonlinear interpolation of
Hamiltonians in Eq. (\ref{hadiabatic}). For more details on the possibility of
considering a nonlinear interpolation when $\left\langle w|s\right\rangle
=\pi/2$, we refer to Ref. \cite{ali04}. The time-dependent function $s\left(
t\right)  $ that defines the linear interpolation is a smooth function such
that $s\left(  0\right)  =0$ and $s\left(  T\right)  =1$, with $T$ being the
run time of the search algorithm. Stated otherwise, the functions $s\left(
t\right)  $ and $1-s\left(  t\right)  $ denote a turn-on and a turn-off
function, respectively. The basic idea in the adiabatic quantum search
algorithm can be expressed as follows. Assuming that the ground state of the
Hamiltonian $\mathcal{H}_{1}$ encodes the solution of the search problem, to
find the target state with large probability, we prepare the system in the
ground state of the Hamiltonian $\mathcal{H}_{0}$. Then, we change
$\mathcal{H}_{0}$ adiabatically to $\mathcal{H}_{1}$ from $t=0$ to $t=T$. The
fulfillment of the adiabaticy condition determines the possible temporal
expressions of the function $s\left(  t\right)  $. In particular, adapting the
evolution rate $ds/dt$ to the local adiabaticity condition%
\begin{equation}
\left\vert \frac{ds}{dt}\right\vert \leq\varepsilon\frac{\left[  \lambda
_{1}\left(  t\right)  -\lambda_{0}\left(  t\right)  \right]  ^{2}%
}{\left\langle \lambda_{1}\left(  t\right)  |\frac{d\mathcal{H}%
_{\text{adiabatic}}}{dt}|\lambda_{0}\left(  t\right)  \right\rangle }\text{,}
\label{local}%
\end{equation}
Roland and Cerf obtained the following optimum expression of $s\left(
t\right)  $ in Ref. \cite{roland02},%
\begin{equation}
s\left(  t\right)  =\frac{1}{2}+\frac{1}{2}\frac{x}{\sqrt{1-x^{2}}}\tan\left[
2\varepsilon x\sqrt{1-x^{2}}t-\tan^{-1}\left(  \frac{\sqrt{1-x^{2}}}%
{x}\right)  \right]  \text{.} \label{cerf}%
\end{equation}
The quantity $\left[  \lambda_{1}\left(  t\right)  -\lambda_{0}\left(
t\right)  \right]  $ in Eq. (\ref{local}) is the energy gap between the ground
state $\left\vert \lambda_{0}\left(  t\right)  \right\rangle $ and the first
excited state $\left\vert \lambda_{1}\left(  t\right)  \right\rangle $ of the
Hamiltonian $\mathcal{H}_{\text{adiabatic}}\left(  t\right)  $ while the
\emph{real} parameter $\varepsilon$ is such that $0<\varepsilon\ll1$. We
remark that the reduced Planck constant $\hslash$ is set equal to one in Eq.
(\ref{local}) so that $\varepsilon$ has the physical dimensions of an energy
with joules equal to seconds$^{-1}$ since $\hslash=1$. Note that imposing
$s\left(  T\right)  =1$ in Eq.(\ref{cerf}), in the limit of $N$ approaching
infinity with $x=1/\sqrt{N}$ and $N$ being the dimensionality of the search
space, we can obtain the typical Grover-like scaling behavior (for instance,
see the first relation in Eq. (\ref{GF})),
\begin{equation}
T\overset{x\ll1}{\approx}\frac{\pi}{2\varepsilon x}\text{.}%
\end{equation}
The method used by Dalzell, Yoder, and Chuang to establish whether or not a
given adiabatic quantum search Hamiltonian possesses the fixed-point property
was developed by exploiting a physics intuition. Specifically, they noted that
the quantum search Hamiltonian in Eq. (\ref{hadiabatic}) can be formally
viewed as a Hamiltonian describing the quantum evolution of a spin-$1/2$
particle (an electron, for instance) in an external magnetic field. Then,
using a Bloch sphere geometric description, the evolution of the quantum state
of the system can be described as a spin precessing on the sphere around the
magnetic field vector at a temporal rate proportional to the magnitude of the
field itself. At the beginning, the source state $\left\vert s\right\rangle $
(that is, the spin of the particle at $t=0$) and the magnetic field vectors
are parallel. At the end of the evolution, while the magnetic field vector is
parallel to the target state $\left\vert w\right\rangle $, the final state
$\left\vert \psi\left(  T\right)  \right\rangle $ at $t=T$ (that is, the spin
of the particle at $t=T$) is only approximately parallel to $\left\vert
w\right\rangle $. If the search algorithm has a fixed-point property, then
there must exist an upper bound on the angle between $\left\vert \psi\left(
T\right)  \right\rangle $ and $\left\vert w\right\rangle $. The existence of
such an upper bound was estimated, in certain scenarios, by Dalzell, Yoder,
and Chuang in Ref. \cite{dalzell17} thanks to the above mentioned geometric
description of physical origin together with clever sequences of
time-dependent change of coordinates to study the quantum mechanical
Schr\"{o}dinger evolution.

We briefly recall that in Grover's original search scheme, the algorithm must
be stopped at a precise instant in order to have a large success probability.
Instead, a search algorithm shows a fixed-point property when the transition
probability from the source state to the target state remains high even when
the algorithm runs longer than necessary. In Ref. \cite{dalzell17}, the
authors consider algorithms specified by schedules $s\left(  t\text{;
}\varepsilon\text{, }w\right)  $ parametrized in terms of two \emph{real}
parameters $\varepsilon$ and $w$. The former parameter quantifies how fast the
interpolation between the two static Hamiltonians occurs, while\textbf{ }the
latter is a lower bound on the fraction of target states $M/N$ with $w\leq
M/N$. Observe that $M$ denotes the number of target states while $N$ is the
dimensionality of the Hilbert search space. The algorithm has a Grover-like
scaling if the run time $T$ is $T\left(  w\right)  =\mathcal{O}\left(
1/\sqrt{w}\right)  $, that is, there exists positive constants $c$ and $w_{0}$
such that
\begin{equation}
\left\vert T\left(  w\right)  \right\vert \leq c/\sqrt{w}\text{,}%
\end{equation}
for any $w\geq w_{0}$. Furthermore, the algorithm has a fixed-point property
if there exists a function $\delta\left(  \varepsilon\right)  $ such that,%
\begin{equation}
\mathcal{P}_{\left\vert s\right\rangle \rightarrow\left\vert w\right\rangle
}\left(  T\text{; }\varepsilon\text{, }w\right)  \geq1-\left[  \delta\left(
\varepsilon\right)  \right]  ^{2}\text{,}%
\end{equation}
for all $w\leq M/N$ and $\delta\left(  \varepsilon\right)  $ approaches zero
as $\varepsilon$ goes to zero.

\begin{table}[t]
\centering
\begin{tabular}
[c]{c|c|c}\hline\hline
Speed of the schedule & Fixed-point property & Grover-like scaling\\\hline
$\varepsilon$ & no & no\\
$\varepsilon w$ & yes & no\\
$\frac{\varepsilon}{\sqrt{w\left(  1-w\right)  }}\Delta_{w}^{3}$ & no & yes\\
$\varepsilon\Delta_{w}^{3}$ & yes & no\\
$\varepsilon\Delta_{w}^{2}$ & yes & yes\\\hline
\end{tabular}
\caption{Fixed-point and Grover-like scaling behaviors for a variety of
adiabatic search algorithms specified by an energy gap $\Delta_{w}=\Delta
_{w}\left(  s\left(  t\text{; }\varepsilon\text{, }w\right)  \right)  $ and a
schedule $s\left(  t\text{; }\varepsilon\text{, }w\right)  $ with speed
$ds/dt$.}%
\end{table}Depending on the choice of the schedule $s\left(  t\text{;
}\varepsilon\text{, }w\right)  $, Dalzell, Yoder, and Chuang were able to find
a variety of types of search algorithms: some algorithms have both Grover-like
scaling and fixed-point property, some have Grover-like scaling but are not
fixed-point, and some are fixed-point but lack Grover-like scaling. For
example, algorithms defined by means of a schedule $s\left(  t\text{;
}\varepsilon\text{, }w\right)  $ with a temporal rate of change $ds/dt$
proportional to the second power of the energy gap $\Delta_{w}\left(
s\right)  \overset{\text{def}}{=}\sqrt{1-4s\left(  1-s\right)  \left(
1-w\right)  }$ like the one in Ref. \cite{roland02} exhibit both a Grover-like
scaling and a fixed-point property. For an overview of the various scenarios
covered in Ref. \cite{dalzell17}, we refer to Table II.

Interestingly, the imposition of the local adiabaticity condition by Roland
and Cerf together with the idea of using a sequence of time-dependent change
of coordinates to study the Schr\"{o}dinger evolution as proposed by Dalzell,
Yoder, and Chuang find their inspiration into the physics of magnetic
resonance phenomena. In the former case, the authors were inspired by the
so-called adiabatic first passage method employed in nuclear magnetic
resonance experiments \cite{rabi38,powles58}. In the latter case, instead, the
authors were influenced by the theoretical methods used to explain the process
of passing through resonance by studying the perturbation in a rotating
coordinate system where it becomes non-resonant \cite{rabi54}.

\subsection{The nonadiabatic framework}

The quantum mechanical evolution in the case of nonadiabatic quantum search
can be described in terms of a time-dependent Hamiltonian $\mathcal{H}%
_{\text{nonadiabatic}}\left(  t\right)  $ as \cite{romanelli07},%
\begin{equation}
\mathcal{H}_{\text{nonadiabatic}}\left(  t\right)  \overset{\text{def}}%
{=}f\left(  t\right)  \mathcal{H}_{0}+g\left(  t\right)  \mathcal{H}%
_{1}\text{,} \label{nonadiabaticH}%
\end{equation}
where $\mathcal{H}_{0}\overset{\text{def}}{=}\mathcal{I-}\left\vert
s\right\rangle \left\langle s\right\vert $, $\mathcal{H}_{1}\overset
{\text{def}}{=}\mathcal{I-}\left\vert w\right\rangle \left\langle w\right\vert
$, and $\left\langle w|s\right\rangle \overset{\text{def}}{=}x\in\left(
0\text{, }1\right)  $. Unlike the adiabatic scenario, it is not required in
this case to adiabatically drive the source state (that is, the known ground
state of $\mathcal{H}_{0}$) into the target state (that is, the unknown ground
state of $\mathcal{H}_{1}$). For this reason, there is a certain freedom in
choosing the expressions of $f\left(  t\right)  $ and $g\left(  t\right)  $
in\ Eq. (\ref{nonadiabaticH}). In Ref. \cite{romanelli07}, Perez and Romanelli
investigated the dynamics generated by two nonadiabatic quantum search
Hamiltonians as in Eq. (\ref{nonadiabaticH}). In the first case, the functions
$f\left(  t\right)  $ and $g\left(  t\right)  $ were chosen to be equal to%
\begin{equation}
f\left(  t\right)  \overset{\text{def}}{=}-\frac{1}{2}\frac{1}{x\sqrt{1-x^{2}%
}}\left\vert \alpha t+\gamma\right\vert \sin\left(  \theta_{0}+2\Omega
_{0}t\right)  \text{,} \label{f1}%
\end{equation}
and,%
\begin{equation}
g\left(  t\right)  \overset{\text{def}}{=}-\frac{1}{2}\frac{1}{x\sqrt{1-x^{2}%
}}\left\vert \alpha t+\gamma\right\vert \left[  2x\sqrt{1-x^{2}}\cos\left(
\theta_{0}+2\Omega_{0}t\right)  +\left(  1-2x^{2}\right)  \sin\left(
\theta_{0}+2\Omega_{0}t\right)  \right]  \text{,} \label{g1}%
\end{equation}
respectively. Note that $\theta\left(  t\right)  \overset{\text{def}}{=}%
\theta_{0}+2\Omega_{0}t$ denotes the mixing angle, $\gamma\overset{\text{def}%
}{=}\Omega_{\mathcal{H}}\left(  0\right)  $, and $\alpha\overset{\text{def}%
}{=}\left(  d\Omega_{\mathcal{H}}/dt\right)  _{t=0}$. In this first case, it
was shown in Ref. \cite{romanelli07} that the transition probability
$\mathcal{P}_{\left\vert s\right\rangle \rightarrow\left\vert w\right\rangle
}\left(  t\right)  $ exhibits an oscillatory and periodic temporal behavior
where the amplitude of these oscillations remains constant in time.
Furthermore, the target state is found at a specific instant in time that
shows the typical Grover-like scaling behavior since it is proportional to
$\sqrt{N}=1/x$. In the second algorithm, the functions $f\left(  t\right)  $
and $g\left(  t\right)  $ were imposed to be equal to%
\begin{equation}
f\left(  t\right)  \overset{\text{def}}{=}1\text{,} \label{f2}%
\end{equation}
and,%
\begin{equation}
g\left(  t\right)  \overset{\text{def}}{=}1-2x^{2}+x\sqrt{1-x^{2}}\left(
b+ct\right)  \text{,} \label{gigi}%
\end{equation}
respectively. The two quantities $b$ and $c$ in Eq. (\ref{gigi}) are two
parameters that characterize the function $g\left(  t\right)  $. In this
second case, the transition probability $\mathcal{P}_{\left\vert
s\right\rangle \rightarrow\left\vert w\right\rangle }\left(  t\right)  $
exhibits an oscillatory but non-periodic behavior where the amplitude of the
oscillations decreases in time. In other words, although in a\textbf{
}non-monotonic fashion, the search Hamiltonian nevertheless drives the system
towards a fixed point and the asymptotic state at time $T\propto\sqrt{N}=1/x$
overlaps with a large probability (although not exactly equal to one) with the
target state. In both scenarios, the precise expressions of the functions
$f\left(  t\right)  $ and $g\left(  t\right)  $ were determined by imposing
that the coupled system of linear ODEs with time-dependent coefficients that
arise from the time-dependent Schr\"{o}dinger equation $i\hslash\partial
_{t}\mathcal{U}\left(  t\right)  =\mathcal{H}\left(  t\right)  \mathcal{U}%
\left(  t\right)  $ leads to analytical solutions. Indeed, the integration
problem reduces in both cases to a Weber-like differential equation whose
general solution is a superposition of parabolic cylinder functions
\cite{watson,irene}. This type of mathematical scheme was originally employed
by Zener in Ref. \cite{zener32a}. For further details on this specific
technical aspect, we refer to Appendix D.

In the next section, in view of the results achieved thus far, we investigate
the possibility of finding suitable magnetic field configurations yielding
transition probabilities with a monotonic temporal behavior that one would
observe in a quantum dynamical evolution under a fixed-point quantum search
Hamiltonian. In addition, we shall discuss the connection between the schedule
of a search algorithm, in both adiabatic and nonadiabatic quantum mechanical
evolutions, and the control fields in a time-dependent driving Hamiltonian.

\section{Monotonic behavior and time-dependent Hamiltonians}

So far, we have assumed to know \emph{a priori} the configuration of the
external time-dependent magnetic field that specifies the external
perturbation of the quantum system. In addition, we have considered quantum
dynamical evolutions causing a periodic oscillatory behavior of the transition
probabilities which was required to single out a specific instant in time in
order to find the target state with certainty. In this section, we depart from
these working conditions in a number of ways. Firstly, since it is challenging
to find the exact analytical solutions to the Schr\"{o}dinger equation, the
exact configuration of the magnetic field will only be determined \emph{a
posteriori}. Secondly, we shall be considering quantum dynamical evolutions
yielding transition probabilities that present a temporal monotonic
convergence towards the target state. Specifically, the asymptotic quantum
state can overlap with the target state with a large probability and no
specific instant in time need to be selected in order to find the desired
state with almost certainty. Thus, thanks to the analogies between the
dynamics generated by a time-independent Grover-like quantum search
Hamiltonian and a time-dependent Rabi-like two-level quantum Hamiltonian
presented in Sections II and III, a comparison between the dynamics of the
newly selected time-dependent Hamiltonians and that of quantum search
Hamiltonians with a fixed-point property is brought forward in this section.
Inspired by Messina and collaborators, we propose two novel applications of
the techniques proposed in Refs. \cite{messina14,grimaudo18} in order to study
two-level systems in an analytical fashion. This way, we aim to uncover
additional relevant physical insights about quantum search and two-level
quantum systems relying on the work presented in Sections II, III, and IV.

\textbf{ }Assume that the matrix representation with respect to the
computational basis $\mathcal{B}_{\text{canonical}}\overset{\text{def}}%
{=}\left\{  \left\vert w\right\rangle \overset{\text{def}}{=}\binom{1}%
{0}\text{, }\left\vert r\right\rangle \overset{\text{def}}{=}\binom{0}%
{1}\right\}  $ of a time-dependent $\emph{su}\left(  2\right)  $-Hamiltonian
$\mathcal{H}\left(  t\right)  $ is given by,%
\begin{equation}
\left[  \mathcal{H}\left(  t\right)  \right]  _{\mathcal{B}_{\text{canonical}%
}}\overset{\text{def}}{=}\left(
\begin{array}
[c]{cc}%
\Omega\left(  t\right)  & \omega\left(  t\right) \\
\omega^{\ast}\left(  t\right)  & -\Omega\left(  t\right)
\end{array}
\right)  \text{,} \label{ham1}%
\end{equation}
where $\omega\left(  t\right)  $ and $\Omega\left(  t\right)  $ in Eq.
(\ref{ham1}) denote the so-called \emph{complex} transverse field and
\emph{real} longitudinal field, respectively. We convey that longitudinal
fields are oriented\textbf{ }along the $z$-direction while transversal fields
lie in the $xy$-plane. Observe that, up to a term proportional to the identity
(see Eqs. (\ref{canonicalbs1}) and (\ref{canonicalbs})), $\left[
\mathcal{H}\left(  t\right)  \right]  _{\mathcal{B}_{\text{canonical}}}$
parametrizes the most general form of a $\left(  2\times2\right)  $-Hermitian
matrix. Moreover, recall that we are interested in computing transition
probabilities which, in turn, are given in terms of the squared modulus of
\emph{complex} probability amplitudes. Therefore, since the term proportional
to the identity would only generate a time-dependent phase factor, it can
simply be ignored. For the sake of completeness, we also point out that the
reason why we used the $\emph{su}\left(  2\right)  $-terminology is because
the matrix representation of the Hamiltonian in Eq. (\ref{ham1}) can be
rewritten in terms of a linear superposition of the three generators $\left\{
i\sigma_{x}\text{, }-i\sigma_{y}\text{, }i\sigma_{z}\right\}  $ of
$\emph{su}\left(  2\right)  $, the Lie algebra of the special unitary group
$\emph{SU}\left(  2\right)  $. Furthermore, the elements of $\emph{su}\left(
2\right)  $ are anti-Hermitian $\left(  2\times2\right)  $- matrices with
trace zero. In particular, in analogy to Eqs. (\ref{fu1}), (\ref{omegarabi1}),
and (\ref{fu2}), we point out that $\mathcal{H}\left(  t\right)  $ with its
matrix representation in Eq. (\ref{ham1}) can be recast as,%
\begin{equation}
\mathcal{H}\left(  t\right)  =\frac{1}{2}\Omega_{\mathcal{H}}\left(  t\right)
\hat{n}\left(  t\right)  \cdot\vec{\sigma}\text{,} \label{angel}%
\end{equation}
with $\Omega_{\mathcal{H}}\left(  t\right)  $ and $\hat{n}\left(  t\right)  $
given by,
\begin{equation}
\Omega_{\mathcal{H}}\left(  t\right)  \overset{\text{def}}{=}2\sqrt{\left\vert
\omega\left(  t\right)  \right\vert ^{2}+\left[  \Omega\left(  t\right)
\right]  ^{2}}\text{,} \label{omegaH}%
\end{equation}
and,%
\begin{equation}
\hat{n}\left(  t\right)  \overset{\text{def}}{=}\frac{1}{\Omega_{\mathcal{H}%
}\left(  t\right)  }\left(  2\left\vert \omega\left(  t\right)  \right\vert
\cos\left[  \phi_{\omega}\left(  t\right)  \right]  \text{, }-2\left\vert
\omega\left(  t\right)  \right\vert \sin\left[  \phi_{\omega}\left(  t\right)
\right]  \text{, }2\Omega\left(  t\right)  \right)  \text{,}%
\end{equation}
respectively. For further technical details that can be helpful to explain in
more detail what follows in this section, we refer to Appendix E.

\subsection{Applications}

In what follows, we consider two quantum dynamical evolution scenarios
specified by suitable time-dependent magnetic field configurations that
generate transition probabilities exhibiting a temporal monotonic convergence
from the source to the target states. In the first scenario, we assume that
the phase $\phi_{\omega}\left(  t\right)  $ and the magnitude $\left\vert
\omega\left(  t\right)  \right\vert $ of the transverse field $\omega\left(
t\right)  $ are interconnected. In the second scenario, although keeping the
same formal expression of $\omega\left(  t\right)  $, we do not assume any
connection between the phase and the intensity of the \emph{complex}
transverse field.

\subsubsection{Interconnection between phase and magnitude of the transverse
field}

In this first scenario, assume that the \emph{complex} auxiliary function
$X\left(  t\right)  $ is given by
\begin{equation}
X\left(  t\right)  \overset{\text{def}}{=}c\sin\left[  \phi\left(  t\right)
\right]  e^{i\phi\left(  t\right)  }\text{,} \label{xte}%
\end{equation}
with $\phi\left(  t\right)  \in%
\mathbb{R}
$, $c\in%
\mathbb{R}
$, and $\phi\left(  0\right)  =0$. Furthermore, assume that the \emph{real}
longitudinal field $\Omega\left(  t\right)  $ is fixed and the \emph{complex}
transverse field $\omega\left(  t\right)  $ is given by,%
\begin{equation}
\omega\left(  t\right)  \overset{\text{def}}{=}\omega_{0}e^{-\xi t}%
e^{i\phi_{\omega}\left(  t\right)  }\text{,} \label{omega2}%
\end{equation}
with known $\omega_{0}>0$ and $\xi>0$ (that is, only $\left\vert \omega\left(
t\right)  \right\vert $ is fixed \emph{a priori}). Use of the condition
$\omega\left(  t\right)  =\alpha^{2}\left(  t\right)  \dot{X}\left(  t\right)
$ and Eq. (\ref{xte}) imply that if one imposes the working assumption that
the \emph{a priori} unknown phase $\phi_{\omega}\left(  t\right)  $ in\ Eq.
(\ref{omega2}) is such that \cite{messina14},%
\begin{equation}
\frac{\left\vert \omega\left(  t\right)  \right\vert }{c}=\frac{\Omega\left(
t\right)  }{\hslash}+\frac{\dot{\phi}_{\omega}\left(  t\right)  }{2}\text{,}
\label{GRC}%
\end{equation}
an exact analytical solution to the Schr\"{o}dinger evolution can be found.
Equation (\ref{GRC}) is the so-called generalized time-dependent out of
resonance condition in a generalized Rabi-like scenario
\cite{messina14,grimaudo18}. Furthermore, having fixed $\left\vert
\omega\left(  t\right)  \right\vert $, $\Omega\left(  t\right)  $, and
$X\left(  t\right)  $, the phase $\phi_{\omega}\left(  t\right)  $ that
appears in Eq. (\ref{omega2}) can be obtained by integrating with respect to
time Eq. (\ref{GRC}). Taking into account these working conditions and
following the analysis reported in the previous subsection, it can readily be
shown that the \emph{complex} probability amplitudes $\alpha\left(  t\right)
$ and $\beta\left(  t\right)  $ are given by%
\begin{equation}
\alpha\left(  t\right)  =\sqrt{\frac{\hslash^{2}+c^{2}\cos^{2}\left[
\Phi\left(  t\right)  \right]  }{\hslash^{2}+c^{2}}}\exp\left\{  i\left[
\frac{\phi_{\omega}\left(  t\right)  }{2}-\tan^{-1}\left(  \frac{\hslash
}{\sqrt{\hslash^{2}+c^{2}}}\tan\left[  \Phi\left(  t\right)  \right]  \right)
\right]  \right\}  \text{,} \label{ayou}%
\end{equation}
and,%
\begin{equation}
\beta\left(  t\right)  =\frac{c}{\sqrt{\hslash^{2}+c^{2}}}\sin\left[
\Phi\left(  t\right)  \right]  \exp\left\{  i\left[  \frac{\phi_{\omega
}\left(  t\right)  }{2}-\frac{\pi}{2}\right]  \right\}  \text{,} \label{byou}%
\end{equation}
respectively. The \emph{real} function $\Phi\left(  t\right)  $ in Eqs.
(\ref{ayou}) and (\ref{byou}) is defined as,%
\begin{equation}
\Phi\left(  t\right)  \overset{\text{def}}{=}\frac{\sqrt{\hslash^{2}+c^{2}}%
}{\hslash c}\int_{0}^{t}\left\vert \omega\left(  t^{\prime}\right)
\right\vert dt^{\prime}\text{.} \label{fim}%
\end{equation}
Using Eqs. (\ref{ayou}), (\ref{byou}), (\ref{fim}) and, in addition, noting
that%
\begin{equation}
\alpha\left(  t\right)  \beta^{\ast}\left(  t\right)  +\alpha^{\ast}\left(
t\right)  \beta\left(  t\right)  =\frac{2\left\vert \alpha\left(  t\right)
\right\vert ^{2}}{\hslash}c\sin\left[  \phi\left(  t\right)  \right]  \text{,}%
\end{equation}
the transition probability $\mathcal{P}_{\left\vert s\right\rangle
\rightarrow\left\vert w\right\rangle }\left(  t\right)  $ in Eq. (\ref{TP1})
becomes,%
\begin{equation}
\mathcal{P}_{\left\vert s\right\rangle \rightarrow\left\vert w\right\rangle
}\left(  t\right)  =\left\vert \alpha\left(  t\right)  \right\vert ^{2}%
x^{2}+\left\vert \beta\left(  t\right)  \right\vert ^{2}\left(  1-x^{2}%
\right)  +\frac{2\left\vert \alpha\left(  t\right)  \right\vert ^{2}}{\hslash
}c\sin\left[  \phi\left(  t\right)  \right]  x\sqrt{1-x^{2}}\text{.}
\label{tp2}%
\end{equation}
The\emph{ real} phase $\phi\left(  t\right)  $ in Eq. (\ref{tp2}) is defined
as,%
\begin{equation}
\phi\left(  t\right)  \overset{\text{def}}{=}\tan^{-1}\left\{  \frac{\hslash
}{\sqrt{\hslash^{2}+c^{2}}}\tan\left[  \Phi\left(  t\right)  \right]
\right\}  \text{.}%
\end{equation}
Using Eqs. (\ref{omega2}), (\ref{ayou}), (\ref{byou}), and (\ref{fim}), the
squared modulus quantities $\left\vert \alpha\left(  t\right)  \right\vert
^{2}$ and $\left\vert \beta\left(  t\right)  \right\vert ^{2}$ become%
\begin{equation}
\left\vert \alpha\left(  t\right)  \right\vert ^{2}=\frac{1}{\hslash^{2}%
+c^{2}}\left\{  \hslash^{2}+c^{2}\cos^{2}\left[  \frac{\sqrt{\hslash^{2}%
+c^{2}}}{\hslash c}\frac{\omega_{0}}{\xi}\left(  1-e^{-\xi t}\right)  \right]
\right\}  \text{,} \label{A1}%
\end{equation}
and,%
\begin{equation}
\left\vert \beta\left(  t\right)  \right\vert ^{2}=\frac{c^{2}}{\hslash
^{2}+c^{2}}\sin^{2}\left[  \frac{\sqrt{\hslash^{2}+c^{2}}}{\hslash c}%
\frac{\omega_{0}}{\xi}\left(  1-e^{-\xi t}\right)  \right]  \text{,}
\label{B1}%
\end{equation}
respectively. Observe that at the resonance (that is, when $c$ approaches
infinity in Eq. (\ref{GRC})) and in the asymptotic limit of time approaching
infinity, $\left\vert \alpha\left(  t\right)  \right\vert ^{2}$ and
$\left\vert \beta\left(  t\right)  \right\vert ^{2}$ approach zero and one,
respectively, provided that the coefficient $\xi$ is defined as%
\begin{equation}
\xi\overset{\text{def}}{=}\frac{2}{\left(  2n+1\right)  \pi}\frac
{\sqrt{\hslash^{2}+c^{2}}}{c}\frac{\omega_{0}}{\hslash}\text{,} \label{csi}%
\end{equation}
where $n\in%
\mathbb{Z}
$. To obtain a monotonic convergent behavior without any transient
oscillation, we impose $n=0$ in Eq. (\ref{csi}). For $n\in%
\mathbb{Z}
\backslash\left\{  0\right\}  $, the convergence is non-monotonic since it
appears only after some transient oscillations. Furthermore, note that
$c\sin\left[  \phi\left(  t\right)  \right]  $ in the expression of
$\mathcal{P}_{\left\vert s\right\rangle \rightarrow\left\vert w\right\rangle
}\left(  t\right)  $ in\ Eq. (\ref{TP1}) is given by
\begin{equation}
c\sin\left[  \phi\left(  t\right)  \right]  =\frac{c\tan\left[  \phi\left(
t\right)  \right]  }{\sqrt{1+\tan^{2}\left[  \phi\left(  t\right)  \right]  }%
}\text{,} \label{csi1}%
\end{equation}
where,%
\begin{equation}
\tan\left[  \phi\left(  t\right)  \right]  =\frac{\hslash}{\sqrt{\hslash
^{2}+c^{2}}}\tan\left[  \frac{\sqrt{\hslash^{2}+c^{2}}}{\hslash c}\frac
{\omega_{0}}{\xi}\left(  1-e^{-\xi t}\right)  \right]  \text{.} \label{csi2}%
\end{equation}
Therefore, using Eqs. (\ref{csi}), (\ref{csi1}) and (\ref{csi2}), in the
asymptotic limit of time approaching infinity at the resonance, we find that
$c\sin\left[  \phi\left(  t\right)  \right]  $ approaches zero.\ 

\begin{figure}[t]
\centering
\includegraphics[width=0.35\textwidth] {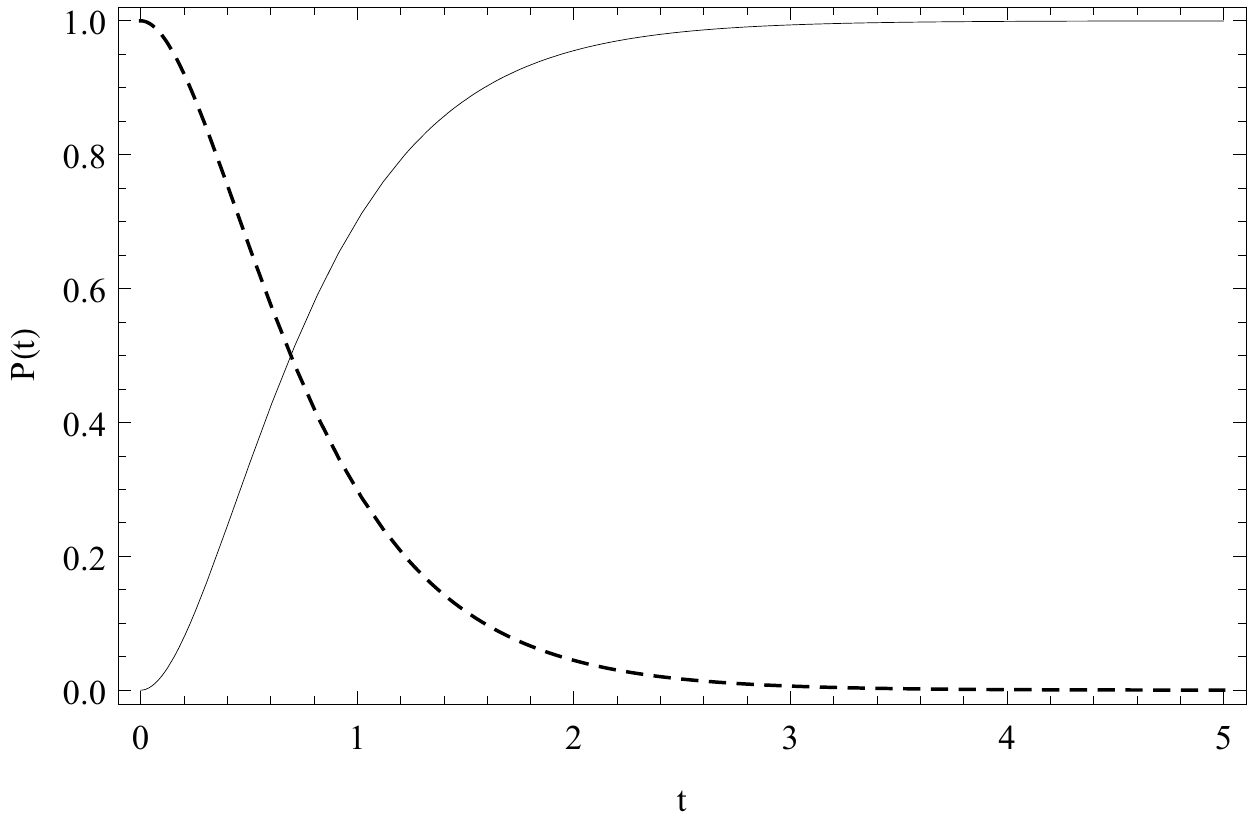}\caption{Monotonic temporal
behavior of the transition probabilities $\mathcal{P}_{\left\vert
s\right\rangle \rightarrow\left\vert w\right\rangle }\left(  t\right)  $
(solid line) and $\mathcal{P}_{\left\vert s\right\rangle \rightarrow\left\vert
r\right\rangle }\left(  t\right)  $ (dashed line) at the generalized Rabi
resonance where $c$ approaches infinity with $x=0$ and $\xi=1$.}%
\label{fig3}%
\end{figure}Finally, we can conclude that $\mathcal{P}_{\left\vert
s\right\rangle \rightarrow\left\vert w\right\rangle }\left(  t\right)  $ in
Eq. (\ref{tp2}) asymptotically approaches the limiting value of $1-x^{2}$
under the above mentioned working conditions. In the additional working
assumption of very small $x$, one can achieve a large numerical value of the
transition probability $\mathcal{P}_{\left\vert s\right\rangle \rightarrow
\left\vert w\right\rangle }$ that, ideally, can approach unity. A plot that
exhibits the monotonic temporal behavior of the transition probabilities
$\mathcal{P}_{\left\vert s\right\rangle \rightarrow\left\vert w\right\rangle
}\left(  t\right)  $ and $\mathcal{P}_{\left\vert s\right\rangle
\rightarrow\left\vert r\right\rangle }\left(  t\right)  $ at the generalized
Rabi resonance where $c$ approaches infinity with $x=0$ and $\xi=1$ appears in
Fig. 3.

\subsubsection{Disconnection between phase and magnitude of the transverse
field}

In this second illustrative example, we assume that the \emph{complex}
auxiliary function is given by
\begin{equation}
X\left(  t\right)  \overset{\text{def}}{=}A\left(  t\right)  e^{i\phi\left(
t\right)  }\text{,} \label{xte2}%
\end{equation}
with $A\left(  t\right)  \in%
\mathbb{R}
$, $\phi\left(  t\right)  \in%
\mathbb{R}
$, and $A\left(  0\right)  =0$. Furthermore, assume that the \emph{complex}
transverse field $\omega\left(  t\right)  $ is fixed and defined as,%
\begin{equation}
\omega\left(  t\right)  =\omega_{0}e^{-\xi t}e^{i\phi_{\omega}\left(
t\right)  }\text{,} \label{omegacsi}%
\end{equation}
with known $\omega_{0}>0$ and $\xi>0$ (that is, both $\left\vert \omega\left(
t\right)  \right\vert $ and $\phi_{\omega}\left(  t\right)  $ are fixed
\emph{a priori}). Unlike the previous application, the phase $\phi_{\omega
}\left(  t\right)  $ can be arbitrary now. In this case, using $\omega\left(
t\right)  =\alpha^{2}\left(  t\right)  \dot{X}\left(  t\right)  $ and Eq.
(\ref{xte2}) together with defining a \emph{real} function $\Theta\left(
t\right)  $ with $\Theta\left(  0\right)  =0$ as%
\begin{equation}
\Theta\left(  t\right)  \overset{\text{def}}{=}\tan^{-1}\left(  \frac{\dot{A}%
}{A}\dot{\phi}\right)  \text{,}%
\end{equation}
it happens to be possible to compute exactly the probability amplitudes
$\alpha\left(  t\right)  $ and $\beta\left(  t\right)  $ provided that the
\emph{real} longitudinal field $\Omega\left(  t\right)  $ satisfies the
following relation \cite{messina14},%
\begin{equation}
\Omega\left(  t\right)  =\frac{\hslash}{2}\left[  \dot{\Theta}\left(
t\right)  -\dot{\phi}_{\omega}\left(  t\right)  \right]  +\left\vert
\omega\left(  t\right)  \right\vert \sin\left[  \Theta\left(  t\right)
\right]  \cot\left[  \frac{2}{\hslash}\int_{0}^{t}\left\vert \omega\left(
t^{\prime}\right)  \right\vert \cos\left[  \Theta\left(  t^{\prime}\right)
\right]  dt^{\prime}\right]  \text{.} \label{longitudinal}%
\end{equation}
In particular, following the line of reasoning reported in the previous
subsection, the \emph{complex} probability amplitudes $\alpha\left(  t\right)
$ and $\beta\left(  t\right)  $ become,%
\begin{equation}
\alpha\left(  t\right)  =\cos\left[  \frac{1}{\hslash}\int_{0}^{t}\left\vert
\omega\left(  t^{\prime}\right)  \right\vert \cos\left[  \Theta\left(
t^{\prime}\right)  \right]  dt^{\prime}\right]  \cdot\exp\left[  -i\int
_{0}^{t}\left(  \frac{\dot{\Theta}\left(  t^{\prime}\right)  -\dot{\phi
}_{\omega}\left(  t^{\prime}\right)  }{2}+\frac{\left\vert \omega\left(
t^{\prime}\right)  \right\vert \sin\left[  \Theta\left(  t^{\prime}\right)
\right]  }{\hslash\sin\left[  \frac{2}{\hslash}\int_{0}^{t^{\prime}}\left\vert
\omega\left(  t^{\prime\prime}\right)  \right\vert \cos\left[  \Theta\left(
t^{\prime\prime}\right)  \right]  dt^{\prime\prime}\right]  }\right)
dt^{\prime}\right]  \text{,} \label{cacchio1}%
\end{equation}
and%
\begin{equation}
\beta\left(  t\right)  =\frac{1}{i\hslash}\alpha\left(  t\right)  A\left(
t\right)  e^{i\phi\left(  t\right)  }\text{,} \label{betuccio}%
\end{equation}
respectively. Furthermore, the quantities $A\left(  t\right)  $ and
$\phi\left(  t\right)  $ in Eq. (\ref{betuccio}) are given by,%
\begin{equation}
A\left(  t\right)  \overset{\text{def}}{=}\hslash\tan\left[  \frac{1}{\hslash
}\int_{0}^{t}\left\vert \omega\left(  t^{\prime}\right)  \right\vert
\cos\left[  \Theta\left(  t^{\prime}\right)  \right]  dt^{\prime}\right]
\text{,} \label{cacchio2}%
\end{equation}
and%
\begin{equation}
\phi\left(  t\right)  \overset{\text{def}}{=}\int_{0}^{t}\frac{2\left\vert
\omega\left(  t^{\prime}\right)  \right\vert }{\hslash}\frac{\sin\left[
\Theta\left(  t^{\prime}\right)  \right]  }{\sin\left[  \frac{2}{\hslash}%
\int_{0}^{t^{\prime}}\left\vert \omega\left(  t^{\prime\prime}\right)
\right\vert \cos\left[  \Theta\left(  t^{\prime\prime}\right)  \right]
dt^{\prime\prime}\right]  }dt^{\prime}\text{,}%
\end{equation}
respectively. In what follows, we assume that $\Theta\left(  t\right)  $ is
defined as%
\begin{equation}
\cos\left[  \Theta\left(  t\right)  \right]  \overset{\text{def}}{=}\cosh
^{-2}\left(  \xi t\right)  \text{,} \label{tetadeff}%
\end{equation}
where the coefficient $\xi$ in Eq. (\ref{tetadeff}) plays the same role
covered by $\xi$ that was originally introduced in\ Eq. (\ref{omegacsi}).
Using Eqs. (\ref{cacchio1}), (\ref{betuccio}), and (\ref{cacchio2}), after
some straightforward but tedious algebra, we determine that the squared
modulus quantities $\left\vert \alpha\left(  t\right)  \right\vert ^{2}$ and
$\left\vert \beta\left(  t\right)  \right\vert ^{2}$ are given by%
\begin{equation}
\left\vert \alpha\left(  t\right)  \right\vert ^{2}=\cos^{2}\left\{
\frac{2\omega_{0}}{\hslash\xi}\left[  \frac{e^{\xi t}}{1+e^{2\xi t}}-\tan
^{-1}\left(  e^{-\xi t}\right)  +\left(  \frac{\pi}{4}-\frac{1}{2}\right)
\right]  \right\}  \text{,} \label{A2}%
\end{equation}
and%
\begin{equation}
\left\vert \beta\left(  t\right)  \right\vert ^{2}=\sin^{2}\left\{
\frac{2\omega_{0}}{\hslash\xi}\left[  \frac{e^{\xi t}}{1+e^{2\xi t}}-\tan
^{-1}\left(  e^{-\xi t}\right)  +\left(  \frac{\pi}{4}-\frac{1}{2}\right)
\right]  \right\}  \text{,} \label{B2}%
\end{equation}
respectively. Observe that in the asymptotic limit of time approaching
infinity, $\left\vert \alpha\left(  t\right)  \right\vert ^{2}$ and
$\left\vert \beta\left(  t\right)  \right\vert ^{2}$ approach zero and one in
a strictly monotonic fashion, respectively, provided that the coefficient
$\xi$ in Eq. (\ref{tetadeff}) is defined as%
\begin{equation}
\xi\overset{\text{def}}{=}\frac{4}{\pi}\frac{\omega_{0}}{\hslash}\left(
\frac{\pi}{4}-\frac{1}{2}\right)  \text{.}%
\end{equation}
Using Eqs. (\ref{cacchio1}), (\ref{betuccio}), (\ref{cacchio2}) and, in
addition, noting that%
\begin{equation}
\alpha\left(  t\right)  \beta^{\ast}\left(  t\right)  +\alpha^{\ast}\left(
t\right)  \beta\left(  t\right)  =\frac{2\left\vert \alpha\left(  t\right)
\right\vert ^{2}}{\hslash}A\left(  t\right)  \sin\left[  \phi\left(  t\right)
\right]  \text{,}%
\end{equation}
the transition probability $\mathcal{P}_{\left\vert s\right\rangle
\rightarrow\left\vert w\right\rangle }\left(  t\right)  $ in Eq. (\ref{TP1})
becomes,%
\begin{equation}
\mathcal{P}_{\left\vert s\right\rangle \rightarrow\left\vert w\right\rangle
}\left(  t\right)  =\left\vert \alpha\left(  t\right)  \right\vert ^{2}%
x^{2}+\left\vert \beta\left(  t\right)  \right\vert ^{2}\left(  1-x^{2}%
\right)  +\frac{2\left\vert \alpha\left(  t\right)  \right\vert ^{2}}{\hslash
}A\left(  t\right)  \sin\left[  \phi\left(  t\right)  \right]  x\sqrt{1-x^{2}%
}\text{.} \label{tp3}%
\end{equation}
Noting that $\left\vert \alpha\left(  t\right)  \right\vert ^{2}A\left(
t\right)  $ approaches zero as time approaches infinity, we conclude that
$\mathcal{P}_{\left\vert s\right\rangle \rightarrow\left\vert w\right\rangle
}\left(  t\right)  $ in Eq. (\ref{tp3}) asymptotically approaches the limiting
value of $1-x^{2}$ under the above mentioned working conditions.

\begin{figure}[t]
\centering
\includegraphics[width=0.35\textwidth] {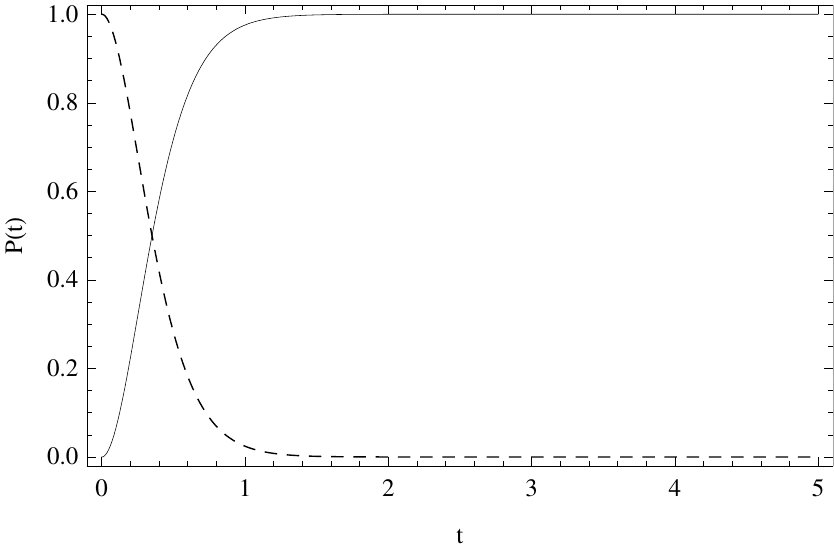}\caption{Monotonic temporal
behavior of the transition probabilities $\mathcal{P}_{\left\vert
s\right\rangle \rightarrow\left\vert w\right\rangle }\left(  t\right)  $
(solid line) and $\mathcal{P}_{\left\vert s\right\rangle \rightarrow\left\vert
r\right\rangle }\left(  t\right)  $ (dashed line) with $x=0$ and $\xi=1$.}%
\label{fig4}%
\end{figure}In analogy to the previous application, under the additional
working assumption of very small $x$, one can achieve a large numerical value
of the transition probability $\mathcal{P}_{\left\vert s\right\rangle
\rightarrow\left\vert w\right\rangle }$ that, ideally, can approach unity. A
plot that illustrates the monotonic temporal behavior of the transition
probabilities $\mathcal{P}_{\left\vert s\right\rangle \rightarrow\left\vert
w\right\rangle }\left(  t\right)  $ and $\mathcal{P}_{\left\vert
s\right\rangle \rightarrow\left\vert r\right\rangle }\left(  t\right)  $ with
$x=0$ and $\xi=1$ appears in Fig. 4. For the sake of completeness, we finally
point out that the explicit expression for the longitudinal field
$\Omega\left(  t\right)  $ in Eq. (\ref{longitudinal}) can be computed by
means of Eqs. (\ref{omegacsi}) and (\ref{tetadeff}) where the phase
$\phi_{\omega}\left(  t\right)  $ can be chosen arbitrarily.

\subsection{Link with adiabatic and nonadiabatic search Hamiltonians}

In the adiabatic case, it is possible to express the schedule $s\left(
t\right)  $ of a search algorithm in terms of the ratio between the transverse
and longitudinal fields $\omega\left(  t\right)  $ and $\Omega\left(
t\right)  $, respectively. Indeed, using Eqs. (\ref{hadiabatic}) and
(\ref{ham1}), it follows that%
\begin{equation}
s\left(  t\right)  =s\left(  \omega\left(  t\right)  \text{, }\Omega\left(
t\right)  \right)  =\frac{2x\sqrt{1-x^{2}}-\frac{\omega\left(  t\right)
}{\Omega\left(  t\right)  }\left(  1-2x^{2}\right)  }{2x\sqrt{1-x^{2}}%
-2\frac{\omega\left(  t\right)  }{\Omega\left(  t\right)  }\left(
1-x^{2}\right)  }\text{.} \label{schedule1}%
\end{equation}
From Eq. (\ref{schedule1}), it is evident that the schedule function $s\left(
t\right)  $ that appears in the Hamiltonian $\mathcal{H}_{\text{adiabatic}%
}\left(  t\right)  $ depends on the ratio between the fields $\omega\left(
t\right)  $ and $\Omega\left(  t\right)  $ when the adiabatic search evolution
is regarded as the quantum mechanical evolution of an electron immersed in a
time-dependent external magnetic field. In particular, the sign of the speed
$ds/dt$ of the schedule $s\left(  t\right)  $ is determined by the rate of
change in time of the ratio $\omega\left(  t\right)  /\Omega\left(  t\right)
$ since%
\begin{equation}
\frac{ds}{dt}=\frac{2x\sqrt{1-x^{2}}}{\left[  2x\sqrt{1-x^{2}}-2\frac
{\omega\left(  t\right)  }{\Omega\left(  t\right)  }\left(  1-x^{2}\right)
\right]  ^{2}}\cdot\frac{d}{dt}\left(  \frac{\omega\left(  t\right)  }%
{\Omega\left(  t\right)  }\right)  \text{.} \label{rate}%
\end{equation}
We point out that in order\textbf{ }to satisfy the boundary conditions
$s\left(  0\right)  =0$ and $s\left(  T\right)  =1$, it is necessary that
$\omega\left(  0\right)  =-x\sqrt{1-x^{2}}$, $\omega\left(  T\right)  =0$,
$\Omega\left(  0\right)  =-\left(  1/2\right)  \left(  1-2x^{2}\right)  $, and
$\Omega\left(  T\right)  =1/2$. For example, in the case of the Roland-Cerf
schedule in Eq. (\ref{cerf}), the transverse field $\omega\left(  t\right)  $
can be written as,%
\begin{equation}
\omega\left(  t\right)  =\frac{1}{2}\left\{  x^{2}\tan\left[  2\varepsilon
x\sqrt{1-x^{2}}t-\tan^{-1}\left(  \frac{\sqrt{1-x^{2}}}{x}\right)  \right]
-x\sqrt{1-x^{2}}\right\}  \text{,}%
\end{equation}
where the run time $T$ equals%
\begin{equation}
T=\frac{1}{\varepsilon}\frac{1}{x\sqrt{1-x^{2}}}\tan^{-1}\left(  \frac
{\sqrt{1-x^{2}}}{x}\right)  \text{.}%
\end{equation}
In the nonadiabatic case, using Eqs. (\ref{nonadiabaticH}) and (\ref{ham1}),
it follows that the analogue of Eq. (\ref{schedule1}) becomes%
\begin{equation}
\frac{\omega\left(  t\right)  }{\Omega\left(  t\right)  }=\frac{2f\left(
t\right)  x\sqrt{1-x^{2}}}{f\left(  t\right)  \left(  1-2x^{2}\right)
-g\left(  t\right)  }\text{,} \label{ration}%
\end{equation}
that is, after some algebra,%
\begin{equation}
g\left(  t\right)  =\left[  \left(  1-2x^{2}\right)  -2\frac{\Omega\left(
t\right)  }{\omega\left(  t\right)  }x\sqrt{1-x^{2}}\right]  f\left(
t\right)  \text{.} \label{gspecial}%
\end{equation}
Interestingly, setting $f\left(  t\right)  =1$, $x=1/\sqrt{N}$, and
$\Omega\left(  t\right)  /\omega\left(  t\right)  =at+b$ with $a$ and $b$ in $%
\mathbb{R}
$, Eq. (\ref{gspecial}) reduces to Eq.(\ref{gigi}), the latter being
introduced without any particular physical motivation by Perez and Romanelli
in Ref. \cite{romanelli07}. From Eq. (\ref{gspecial}), it becomes clear that
once the function $f\left(  t\right)  $ is chosen to be constant, in order to
have the simplest time-dependent nonadiabatic search Hamiltonian
$\mathcal{H}_{\text{nonadiabatic}}\left(  t\right)  $, the function $g\left(
t\right)  $ must depend linearly on time. For the sake of completeness, we
emphasize that the expressions of $f\left(  t\right)  $ and $g\left(
t\right)  $ in Eqs. (\ref{f1}) and (\ref{f2}), respectively, were chosen by
Perez and Romanelli by imposing that $\Omega_{\mathcal{H}}\left(  t\right)  $
in Eq.(\ref{omegaH}) was linear in time. As a final remark, we point out that
if we set $f\left(  t\right)  +g\left(  t\right)  $ equal to zero, from Eq.
(\ref{ration}) it is evident\textbf{ }that the ratio $\omega\left(  t\right)
/\Omega\left(  t\right)  $ becomes constant and equal to%
\begin{equation}
\frac{\omega\left(  t\right)  }{\Omega\left(  t\right)  }=\frac{x}%
{\sqrt{1-x^{2}}}\text{.} \label{inter}%
\end{equation}
Furthermore, requiring that the field $\omega\left(  t\right)  $ be
\emph{real} so that its phase $\phi_{\omega}\left(  t\right)  $ is equal to
zero in Eq. (\ref{GRC}), the generalized out of resonance condition becomes%
\begin{equation}
\frac{\omega\left(  t\right)  }{\Omega\left(  t\right)  }=\frac{c}{\hslash
}\text{.} \label{inter2}%
\end{equation}
Finally, being within the framework of nonadiabatic quantum search, we are
able to find an interpretation of the parameter $c$ that originally appears in
Eq. (\ref{xte}) in terms of the quantum mechanical overlap $x\overset
{\text{def}}{=}\left\langle w|x\right\rangle $ as $c\left(  x\right)  =\hslash
x/\sqrt{1-x^{2}}$ by making use of Eqs. (\ref{inter}) and (\ref{inter2}). In
conclusion, our application of the mathematical methods introduced in Refs.
\cite{messina14,grimaudo18} to quantum search problems is useful for a number
of reasons. First, it helps in\textbf{ }finding search Hamiltonians yielding
exact analytical expressions of transition probabilities from a source state
to a target state with strictly monotonic behavior as evident from Eqs.
(\ref{A1}), (\ref{B1}), (\ref{A2}), and (\ref{B2}).\ Second, it enhances the
physical interpretation of the schedule functions as is evident from Eqs.
(\ref{schedule1}) and (\ref{gspecial}). Finally, it helps in clarifying the
meaning of mathematically introduced quantities, such as the parameter $c$, by
linking them to geometrical quantities with a clear physical significance such
as the quantum mechanical overlap $x$.

\section{Final remarks}

In this paper, we investigated the connection between analog quantum search
and two-level quantum systems. A number of quantitative results together with
some insightful remarks were found. Our main findings can be outlined in some
detail as follows.

\begin{enumerate}
\item First, we characterized in a quantitative manner the intuitive analogy
between a Grover-like quantum search dynamics defined in terms of the
time-independent Hamiltonian in Eq. (\ref{hamilton}) and a Rabi-like quantum
mechanical evolution of a spin-$1/2$ particle in an external time-dependent
magnetic field specified by the time-dependent Hamiltonian in Eq.
(\ref{rabi1}). This task was achieved in two steps. In the first step, we
presented a detailed and exact analytical computation of the transition
probabilities from a source state to a target state in both scenarios as
reported in Eqs. (\ref{it2}) (for details, see Appendix A) and (\ref{P2}) (for
details, see Appendix C). A plot exhibiting the characteristic periodic
oscillatory behavior of the transition probabilities in the Rabi-like case
appears in Fig. 1. In the second step, we provided a schematic correspondence
between relevant physical quantities in the two scenarios, including the
energy gaps and the interaction strengths. Such a correspondence appears in
Table I and, in particular, enabled us to physically interpret the quantum
mechanical overlap between the source state and the target state in the
Grover-like scenario in terms of the ratio of magnetic field intensities in
the Rabi-like framework as evident from Eq. (\ref{chichi2}) and Fig. 2. This
first set of results, motivated by the insightful remarks appearing in Refs.
\cite{dalzell17,byrnes18}, are especially interesting since they quantify for
the first time in a detailed and exact analytical fashion the physical link
between Grover-like quantum algorithms and Rabi oscillations originating from
a time-dependent two-level quantum system Hamiltonian.

\item Second, to further advance our understanding of this physical link
between quantum search problems and two-level quantum systems dynamics
yielding transition probabilities with periodic oscillatory temporal behavior,
we explored the possibility of extending such a physical connection to
fixed-point quantum algorithms specified by a monotonic convergence towards
the target state. To accomplish this task, we first revisited the concepts of
both adiabatic and nonadiabatic quantum search algorithms characterized,
unlike the original time-independent Grover-like search Hamiltonians, by
time-dependent search Hamiltonians. In the adiabatic case, we focused
primarily on the role played by the rate of change in time of the
schedule\textbf{ }$s\left(  t\right)  $ of the algorithm in determining its
Grover-like scaling behavior and/or its fixed-point property feature. A
summary of this specific review appears in Table II. In the nonadiabatic case,
instead, we attempted to determine the motivation behind the choice of
particular temporal expressions of the search Hamiltonians as in Eq.
(\ref{nonadiabaticH}) in terms of their schedule functions that appear in Eqs.
(\ref{f1}), (\ref{g1}), (\ref{f2}), and (\ref{gigi}). Our Section IV
constitutes an original and unifying critical reconsideration of both
adiabatic and nonadiabatic quantum search Hamiltonians as originally presented
in Refs. \cite{dalzell17,romanelli07} in terms of the schedule of the algorithm.

\item Third, we exploited recent mathematical techniques developed within the
framework of exactly solvable two-level time-dependent $\emph{su}\left(
2\right)  $-Hamiltonian models in order to provide two exact analytical
expressions of transition probabilities exhibiting monotonic convergence from
the source state to the target state. The outcomes of our first application
appear in Eqs. (\ref{A1}) and (\ref{B1}) and are displayed in Fig. 3.
Furthermore, the findings of our second application are reported in Eqs.
(\ref{A2}) and (\ref{B2}) and are illustrated in Fig. 4. More interestingly,
thanks to the reformulation of the adiabatic and nonadiabatic search
Hamiltonians in Eqs. (\ref{hadiabatic}) and (\ref{nonadiabaticH}),
respectively, in terms of the $\emph{su}\left(  2\right)  $-Hamiltonian model
in Eq. (\ref{ham1}), we uncovered some relevant insights. In the adiabatic
case, we were able to express the schedule and its speed in terms of the ratio
of the transverse and longitudinal fields that define the particular
time-dependent magnetic field configuration in which the spin-$1/2$ particle
is immersed. These results appear in Eqs. (\ref{schedule1}) and (\ref{rate}).
In the nonadiabatic case, we found a compact general relation between the two
schedule functions $f\left(  t\right)  $ and $g\left(  t\right)  $ where,
again, the ratio between the longitudinal and the transverse fields emerges.
This relation appears in Eq. (\ref{gspecial}) and, interestingly, coincides
with the physically unmotivated choice that appears in Eq. (\ref{gigi}).
Finally, we were able to assign some clear interpretation in terms of
geometric quantities with physical meaning to certain auxiliary mathematical
parameters that enter the above mentioned mathematical techniques. A
particular case of such an assignment appears by combining Eqs. (\ref{inter})
and (\ref{inter2}). The importance of our third set of results is threefold.
First, we extend the application of recently introduced mathematical
techniques in Refs. \cite{messina14,grimaudo18} to quantum search
problems.\ Second, we were\textbf{ }able to give a physical interpretation of
the schedule function in terms of control magnetic fields. Finally, we
extended the physical interpretation of fixed-point quantum search algorithms
in terms of the non-relativistic quantum mechanical evolution of an electron
immersed in specific time-dependent magnetic field configuration.
\end{enumerate}

The main physical insights that emerge from our investigation can be
summarized as follows. A quantum search Hamiltonian can be regarded as a
superposition of an unperturbed Hamiltonian and a perturbation (see Section
III). In particular, the minimum search time appears to be inversely
proportional the energy level separation defined in terms of the eigenvalues
of the (total) Hamiltonian. In principle, by modifying the perturbation, one
can change the minimum search time. The perturbation can be time-independent
or time-dependent. In the case of a time-independent perturbation (see Section
II), the constant perturbation can be characterized by a number of parameters.
By suitably tuning these parameters, one can minimize the minimum search time
or, alternatively, minimize the minimum time to achieve a nearly optimal
success probability value. The same reasoning can be extended to the
time-dependent scenario where, however, a richer realm of possibilities can be
explored. For instance, one can observe relevant dynamical features of a
quantum system by varying the external perturbation in an adiabatic or
nonadiabatic fashion (see Section IV). The role played by the time-dependent
schedules of the search algorithms in both adiabatic and nonadiabatic settings
can be described in terms of a time-dependent perturbation. Finally,
expressing the perturbation in terms of time-dependent transverse and
longitudinal fields, one can\textbf{ }suitably choose one of these two fields
(and, unfortunately, determine the other one only \emph{a posteriori}) in such
a manner that a desired temporal behavior of the transition probability from
the source state to the target state is recovered (see Section V).

Despite our serious effort to propose a unifying perspective on quantum search
algorithms in terms of the physics of two-state quantum systems, our work is
not free of limitations. For instance, a clear limitation of our analysis is
that we were unable to predict the exact or, for that matter, guess an
approximate analytical temporal behavior of the transition probability from
the source state to the target state for an\textbf{ }\emph{a priori}\textbf{
}chosen time-dependent magnetic field configuration. However, the insights
that emerge from our work open up the possibility of exploring a number of
intriguing questions. For instance, given our enhanced understanding of the
role played by the schedule of a search algorithm, we have reason to believe
that the work presented in this paper will help further enhance the
comprehension of our recent information geometric analysis of quantum
algorithms viewed from a statistical thermodynamics standpoint as proposed in
Ref. \cite{paul18}. More generally, given that the phenomenon of anticrossing
is controlled by the parameters behind the perturbation and results into
lowering of energy, it would be worthwhile exploring in a more detailed manner
the link between minimum search time and avoided crossing in the hope of
finding out an optimal tradeoff between speed (minimum search time), fidelity
(success probability), and energy minimization requirements in actual physical
implementations of quantum search algorithms \cite{campbell17}. We leave these
intriguing investigations to future efforts.

\begin{acknowledgments}
C. C. is grateful to the United States Air Force Research Laboratory (AFRL)
Summer Faculty Fellowship Program for providing support for this work. Any
opinions, findings and conclusions or recommendations expressed in this
manuscript are those of the authors and do not necessarily reflect the views
of AFRL.\textbf{ }Finally, constructive criticism from an anonymous referee
leading to an improved version of this manuscript are sincerely acknowledged
by the authors.
\end{acknowledgments}

\appendix

\section{Derivation of Eq. (\ref{it2})}

In this Appendix, we derive Eq. (\ref{it2}) that appears in Section II. Note
that the matrix representation of the Hamiltonian $\mathcal{H}_{\text{GQS}}$
in Eq. (\ref{hamilton}) with respect to the orthonormal basis $\left\{
\left\vert w\right\rangle \text{, }\left\vert r\right\rangle \right\}  $ where
$\left\langle w|r\right\rangle =\delta_{wr}$, with $\delta_{wr}$ denoting the
Kronecker delta, is given by%
\begin{equation}
\left[  \mathcal{H}_{\text{GQS}}\right]  _{\left\{  \left\vert w\right\rangle
\text{, }\left\vert r\right\rangle \right\}  }\overset{\text{def}}{=}\left(
\begin{array}
[c]{cc}%
\left\langle w|\mathcal{H}_{\text{GQS}}|w\right\rangle  & \left\langle
w|\mathcal{H}_{\text{GQS}}|r\right\rangle \\
\left\langle r|\mathcal{H}_{\text{GQS}}|w\right\rangle  & \left\langle
r|\mathcal{H}_{\text{GQS}}|r\right\rangle
\end{array}
\right)  \text{.}%
\end{equation}
Using the above mentioned orthonormality conditions $\left\langle
w|r\right\rangle =\delta_{wr}$ together with Eqs. (\ref{hamilton}) and
(\ref{sr}), we obtain%
\begin{equation}
\left[  \mathcal{H}_{\text{GQS}}\right]  _{\left\{  \left\vert w\right\rangle
\text{, }\left\vert r\right\rangle \right\}  }=\left(
\begin{array}
[c]{cc}%
h_{11} & h_{12}\\
h_{21} & h_{22}%
\end{array}
\right)  \text{,} \label{symm}%
\end{equation}
where,%
\begin{align}
&  h_{11}\overset{\text{def}}{=}E\left[  \alpha+\left(  \beta+\gamma\right)
x+\delta x^{2}\right]  \text{, }h_{12}\overset{\text{def}}{=}E\sqrt{1-x^{2}%
}\left(  \beta+\delta x\right)  \text{,}\nonumber\\
& \nonumber\\
&  h_{21}\overset{\text{def}}{=}E\sqrt{1-x^{2}}\left(  \gamma+\delta x\right)
\text{, }h_{22}\overset{\text{def}}{=}E\delta\left(  1-x^{2}\right)  \text{.}%
\end{align}
Observe that $\mathcal{H}_{\text{GQS}}$ is an Hermitian operator. Therefore,
imposing the condition that $\mathcal{H}_{\text{GQS}}=\mathcal{H}_{\text{GQS}%
}^{\dagger}$ where the dagger symbol \textquotedblleft$\dagger$%
\textquotedblright\ denotes the usual Hermitian conjugation operation,
\begin{equation}
\left(
\begin{array}
[c]{cc}%
h_{11} & h_{12}\\
h_{21} & h_{22}%
\end{array}
\right)  =\left(
\begin{array}
[c]{cc}%
h_{11}^{\ast} & h_{21}^{\ast}\\
h_{12}^{\ast} & h_{22}^{\ast}%
\end{array}
\right)  \text{,} \label{cc}%
\end{equation}
it follows that $\alpha$ and $\delta$ must be \emph{real} coefficients while
$\beta=\gamma^{\ast}$. The symbol \textquotedblleft$\ast$\textquotedblright%
\ in Eq. (\ref{cc}) denotes the usual \emph{complex} conjugation operation.
Let us proceed to diagonalize the Hermitian matrix $\left[  \mathcal{H}%
_{\text{GQS}}\right]  _{\left\{  \left\vert w\right\rangle \text{, }\left\vert
r\right\rangle \right\}  }$ in Eq. (\ref{symm}). The two \emph{real}
eigenvalues $\lambda_{\pm}$ of the matrix are given by,
\begin{equation}
\lambda_{\pm}\overset{\text{def}}{=}\frac{1}{2}\left[  \left(  h_{11}%
+h_{22}\right)  \pm\sqrt{\left(  h_{11}-h_{22}\right)  ^{2}+4h_{12}h_{21}%
}\right]  \text{.} \label{eigen}%
\end{equation}
The eigenspaces $\mathcal{E}_{\lambda_{-}}$ and $\mathcal{E}_{\lambda_{+}}$
that correspond to the eigenvalues $\lambda_{-}$ and $\lambda_{+}$ are given
by%
\begin{equation}
\mathcal{E}_{\lambda_{-}}\overset{\text{def}}{=}\text{Span}\left\{  \left\vert
v_{\lambda_{-}}\right\rangle \right\}  \text{, and }\mathcal{E}_{\lambda_{+}%
}\overset{\text{def}}{=}\text{Span}\left\{  \left\vert v_{\lambda_{+}%
}\right\rangle \right\}  \text{,}%
\end{equation}
respectively. The two eigenvectors $\left\vert v_{\lambda_{-}}\right\rangle $
and $\left\vert v_{\lambda_{+}}\right\rangle $ corresponding to $\lambda_{+}$
and $\lambda_{-}$ can be written as%
\begin{equation}
\left\vert v_{\lambda_{+}}\right\rangle \overset{\text{def}}{=}\left(
\begin{array}
[c]{c}%
\frac{1}{2h_{21}}\left[  \left(  h_{11}-h_{22}\right)  +\sqrt{\left(
h_{11}-h_{22}\right)  ^{2}+4h_{12}h_{21}}\right] \\
1
\end{array}
\right)  \text{,} \label{v1}%
\end{equation}
and,%
\begin{equation}
\left\vert v_{\lambda_{-}}\right\rangle \overset{\text{def}}{=}\left(
\begin{array}
[c]{c}%
\frac{1}{2h_{21}}\left[  \left(  h_{11}-h_{22}\right)  -\sqrt{\left(
h_{11}-h_{22}\right)  ^{2}+4h_{12}h_{21}}\right] \\
1
\end{array}
\right)  \text{,} \label{v2}%
\end{equation}
respectively. For ease of notational simplicity, let us define two
\emph{complex} quantities $A$ and $B$ as follows%
\begin{equation}
A\overset{\text{def}}{=}\frac{1}{2h_{21}}\left[  \left(  h_{11}-h_{22}\right)
-\sqrt{\left(  h_{11}-h_{22}\right)  ^{2}+4h_{12}h_{21}}\right]  \text{,}
\label{anew}%
\end{equation}
and,%
\begin{equation}
B\overset{\text{def}}{=}\frac{1}{2h_{21}}\left[  \left(  h_{11}-h_{22}\right)
+\sqrt{\left(  h_{11}-h_{22}\right)  ^{2}+4h_{12}h_{21}}\right]  \text{.}
\label{bnew}%
\end{equation}
Employing Eqs. (\ref{v1}), (\ref{v2}), (\ref{anew}), and (\ref{bnew}), the
eigenvector matrix $\mathcal{M}_{\mathcal{H}_{\text{GQS}}}$ and its inverse
$\mathcal{M}_{\mathcal{H}_{\text{GQS}}}^{-1}$ corresponding to the Hamiltonian
matrix $\left[  \mathcal{H}_{\text{GQS}}\right]  _{\left\{  \left\vert
w\right\rangle \text{, }\left\vert r\right\rangle \right\}  }$ are given by,%
\begin{equation}
\mathcal{M}_{\mathcal{H}_{\text{GQS}}}\overset{\text{def}}{=}\left(
\begin{array}
[c]{cc}%
A & B\\
1 & 1
\end{array}
\right)  \text{,} \label{mmatrix2}%
\end{equation}
and,%
\begin{equation}
\mathcal{M}_{\mathcal{H}_{\text{GQS}}}^{-1}\overset{\text{def}}{=}\frac
{1}{A-B}\left(
\begin{array}
[c]{cc}%
1 & -B\\
-1 & A
\end{array}
\right)  =\mathcal{M}_{\mathcal{H}_{\text{GQS}}}^{\dagger}\text{,}
\label{mimatrix2}%
\end{equation}
respectively. In terms of the matrices $\mathcal{M}_{\mathcal{H}_{\text{GQS}}%
}$, $\mathcal{M}_{\mathcal{H}_{\text{GQS}}}^{-1}$, and a diagonal matrix
$H_{\text{GQS-diagonal}}$, the matrix $\left[  \mathcal{H}_{\text{GQS}%
}\right]  _{\left\{  \left\vert w\right\rangle \text{, }\left\vert
r\right\rangle \right\}  }$ in\ Eq. (\ref{symm}) can be expressed as%
\begin{equation}
\left[  \mathcal{H}_{\text{GQS}}\right]  _{\left\{  \left\vert w\right\rangle
\text{, }\left\vert r\right\rangle \right\}  }=\mathcal{M}_{\mathcal{H}%
_{\text{GQS}}}H_{\text{GQS-diagonal}}\mathcal{M}_{\mathcal{H}_{\text{GQS}}%
}^{-1}=\left(
\begin{array}
[c]{cc}%
A & B\\
1 & 1
\end{array}
\right)  \left(
\begin{array}
[c]{cc}%
\lambda_{-} & 0\\
0 & \lambda_{+}%
\end{array}
\right)  \left(
\begin{array}
[c]{cc}%
\frac{1}{A-B} & \frac{-B}{A-B}\\
\frac{-1}{A-B} & \frac{A}{A-B}%
\end{array}
\right)  \text{,}%
\end{equation}
where $H_{\text{GQS-diagonal}}$ is given by,%
\begin{equation}
H_{\text{GQS-diagonal}}\overset{\text{def}}{=}\left[  \mathcal{H}_{\text{GQS}%
}\right]  _{\left\{  \left\vert v_{\lambda_{-}}\right\rangle \text{,
}\left\vert v_{\lambda_{+}}\right\rangle \right\}  }=\left(
\begin{array}
[c]{cc}%
\left\langle v_{\lambda-}|\mathcal{H}_{\text{GQS}}|v_{\lambda-}\right\rangle
& \left\langle v_{\lambda-}|\mathcal{H}_{\text{GQS}}|v_{\lambda_{+}%
}\right\rangle \\
\left\langle v_{\lambda_{+}}|\mathcal{H}_{\text{GQS}}|v_{\lambda
-}\right\rangle  & \left\langle v_{\lambda_{+}}|\mathcal{H}_{\text{GQS}%
}|v_{\lambda_{+}}\right\rangle
\end{array}
\right)  =\left(
\begin{array}
[c]{cc}%
\lambda_{-} & 0\\
0 & \lambda_{+}%
\end{array}
\right)  \text{.} \label{hdiagonal}%
\end{equation}
The eigenvalues in\ Eq. (\ref{hdiagonal}) are defined in\ Eq. (\ref{eigen})
while $A$ and $B$ are given in\ Eqs. (\ref{anew}) and (\ref{bnew}),
respectively. At this juncture, we recall that our goal is to compute the time
$t^{\ast}$ such that $\mathcal{P}_{\left\vert s\right\rangle \rightarrow
\left\vert w\right\rangle }\left(  t^{\ast}\right)  =\mathcal{P}_{\max}$ where
the transition probability $\mathcal{P}_{\left\vert s\right\rangle
\rightarrow\left\vert w\right\rangle }\left(  t\right)  $ is defined in Eq.
(\ref{fidelity}). By means of standard matrix algebra methods, $\mathcal{P}%
_{\left\vert s\right\rangle \rightarrow\left\vert w\right\rangle }\left(
t\right)  $ can be recast as%
\begin{equation}
\mathcal{P}_{\left\vert s\right\rangle \rightarrow\left\vert w\right\rangle
}\left(  t\right)  \overset{\text{def}}{=}\left\vert \left\langle
w|e^{-\frac{i}{\hslash}\mathcal{H}_{\text{GQS}}t}|s\right\rangle \right\vert
^{2}=\left\vert \left\langle w|\mathcal{M}_{\mathcal{H}_{\text{GQS}}}%
e^{-\frac{i}{\hslash}H_{\text{GQS-diagonal}}}\mathcal{M}_{\mathcal{H}%
_{\text{GQS}}}^{\dagger}|s\right\rangle \right\vert ^{2}\text{.} \label{pt3}%
\end{equation}
Using the matrix notation with components expressed relative to the
orthonormal basis $\left\{  \left\vert w\right\rangle \text{, }\left\vert
r\right\rangle \right\}  $, states $\left\vert w\right\rangle $ and
$\left\vert s\right\rangle $ are given by
\begin{equation}
\left\vert w\right\rangle \overset{\text{def}}{=}\left(
\begin{array}
[c]{c}%
1\\
0
\end{array}
\right)  \text{, and }\overset{\text{def}}{\left\vert s\right\rangle =}\left(
\begin{array}
[c]{c}%
x\\
\sqrt{1-x^{2}}%
\end{array}
\right)  \text{,} \label{matic2}%
\end{equation}
respectively. Using Eqs. (\ref{mmatrix2}), (\ref{mimatrix2}), and
(\ref{matic2}), the quantum state amplitude $\left\langle w|e^{-\frac
{i}{\hslash}\mathcal{H}_{\text{GQS}}t}|s\right\rangle $ that appears in the
expression of the fidelity $\mathcal{P}_{\left\vert s\right\rangle
\rightarrow\left\vert w\right\rangle }\left(  t\right)  $ in\ Eq. (\ref{pt3})
becomes%
\begin{equation}
\left\langle w|e^{-\frac{i}{\hslash}\mathcal{H}_{\text{GQS}}t}|s\right\rangle
=\frac{1}{A-B}\left[  Ae^{-\frac{i}{\hslash}\lambda_{-}t}\left(
x-B\sqrt{1-x^{2}}\right)  -Be^{-\frac{i}{\hslash}\lambda_{+}t}\left(
x-A\sqrt{1-x^{2}}\right)  \right]  \text{,} \label{part1b}%
\end{equation}
and, consequently, its complex conjugate $\left\langle w|e^{-\frac{i}{\hslash
}\mathcal{H}_{\text{GQS}}t}|s\right\rangle ^{\ast}$ is given by,
\begin{equation}
\left\langle w|e^{-\frac{i}{\hslash}\mathcal{H}_{\text{GQS}}t}|s\right\rangle
^{\ast}=\frac{1}{A-B}\left[  Ae^{\frac{i}{\hslash}\lambda_{-}t}\left(
x-B\sqrt{1-x^{2}}\right)  -Be^{\frac{i}{\hslash}\lambda_{+}t}\left(
x-A\sqrt{1-x^{2}}\right)  \right]  \text{.} \label{part2b}%
\end{equation}
Note that,%
\begin{equation}
e^{-\frac{i}{\hslash}\lambda_{-}t}=e^{-\frac{i}{\hslash}\frac{h_{11}+h_{22}%
}{2}t}e^{i\frac{a}{\hslash}t}\text{ and, }e^{-\frac{i}{\hslash}\lambda_{+}%
t}=e^{-\frac{i}{\hslash}\frac{h_{11}+h_{22}}{2}t}e^{-i\frac{a}{\hslash}%
t}\text{.} \label{aeq}%
\end{equation}
Recalling Eq. (\ref{eigen}), the \emph{real} quantity $a$ in Eq. (\ref{aeq})
is defined as%
\begin{equation}
a\overset{\text{def}}{=}\frac{1}{2}\sqrt{\left(  h_{11}-h_{22}\right)
^{2}+4h_{12}h_{21}}\text{.} \label{adef}%
\end{equation}
Using Eq. (\ref{aeq}), the \emph{complex} probability amplitudes in Eqs.
(\ref{part1b}) and (\ref{part2b}) become%
\begin{equation}
\left\langle w|e^{-\frac{i}{\hslash}\mathcal{H}_{\text{GQS}}t}|s\right\rangle
=e^{-\frac{i}{\hslash}\frac{h_{11}+h_{22}}{2}t}\left[  \frac{A\left(
x-B\sqrt{1-x^{2}}\right)  }{A-B}e^{i\frac{a}{\hslash}t}-\frac{B\left(
x-A\sqrt{1-x^{2}}\right)  }{A-B}e^{-i\frac{a}{\hslash}t}\right]  \text{,}
\label{part3}%
\end{equation}
and,%
\begin{equation}
\left\langle w|e^{-\frac{i}{\hslash}\mathcal{H}_{\text{GQS}}t}|s\right\rangle
^{\ast}=e^{\frac{i}{\hslash}\frac{h_{11}+h_{22}}{2}t}\left[  \frac{A^{\ast
}\left(  x-B^{\ast}\sqrt{1-x^{2}}\right)  }{A^{\ast}-B^{\ast}}e^{-i\frac
{a}{\hslash}t}-\frac{B^{\ast}\left(  x-A^{\ast}\sqrt{1-x^{2}}\right)
}{A^{\ast}-B^{\ast}}e^{i\frac{a}{\hslash}t}\right]  \text{,} \label{part4}%
\end{equation}
respectively. Using Eqs. (\ref{part3})\ and (\ref{part4}) and introducing the
following three quantities%
\begin{equation}
\tilde{A}\overset{\text{def}}{=}\frac{A\left(  x-B\sqrt{1-x^{2}}\right)
}{A-B}\text{, }\tilde{B}\overset{\text{def}}{=}-\frac{B\left(  x-A\sqrt
{1-x^{2}}\right)  }{A-B}\text{, and }\tilde{\alpha}=\frac{a}{\hslash}t\text{,}
\label{newroba}%
\end{equation}
the transition probability $\mathcal{P}_{\left\vert s\right\rangle
\rightarrow\left\vert w\right\rangle }\left(  t\right)  $ in Eq.
(\ref{fidelity}) becomes%
\begin{equation}
\mathcal{P}_{\left\vert s\right\rangle \rightarrow\left\vert w\right\rangle
}\left(  \tilde{\alpha}\right)  =\left[  \tilde{A}e^{i\tilde{\alpha}}%
+\tilde{B}e^{-i\tilde{\alpha}}\right]  \left[  \tilde{A}^{\ast}e^{-i\tilde
{\alpha}}+\tilde{B}^{\ast}e^{i\tilde{\alpha}}\right]  =\left\vert \tilde
{A}\right\vert ^{2}+\left\vert \tilde{B}\right\vert ^{2}+2\tilde{A}\tilde
{B}^{\ast}\cos\left(  2\tilde{\alpha}\right)  \text{,}%
\end{equation}
where we point out that $\tilde{A}\tilde{B}^{\ast}$ is \emph{real} since
$h_{12}=h_{21}^{\ast}$. By exploiting standard trigonometric relations in a
clever sequential order, we obtain%
\begin{align}
\mathcal{P}_{\left\vert s\right\rangle \rightarrow\left\vert w\right\rangle
}\left(  \tilde{\alpha}\right)   &  =\left\vert \tilde{A}\right\vert
^{2}+\left\vert \tilde{B}\right\vert ^{2}+2\tilde{A}\tilde{B}^{\ast}%
\cos\left(  2\tilde{\alpha}\right) \nonumber\\
&  =\left\vert \tilde{A}\right\vert ^{2}+\left\vert \tilde{B}\right\vert
^{2}+2\tilde{A}\tilde{B}^{\ast}\left[  \cos^{2}\left(  \tilde{\alpha}\right)
-\sin^{2}\left(  \tilde{\alpha}\right)  \right] \nonumber\\
&  =\left\vert \tilde{A}\right\vert ^{2}+\left\vert \tilde{B}\right\vert
^{2}+2\tilde{A}\tilde{B}^{\ast}\cos^{2}\left(  \tilde{\alpha}\right)
-2\tilde{A}\tilde{B}^{\ast}\sin^{2}\left(  \tilde{\alpha}\right) \nonumber\\
&  =\left\vert \tilde{A}\right\vert ^{2}\sin^{2}\left(  \tilde{\alpha}\right)
+\left\vert \tilde{A}\right\vert ^{2}\cos^{2}\left(  \tilde{\alpha}\right)
+\left\vert \tilde{B}\right\vert ^{2}\sin^{2}\left(  \tilde{\alpha}\right)
+\left\vert \tilde{B}\right\vert ^{2}\cos^{2}\left(  \tilde{\alpha}\right)
+2\tilde{A}\tilde{B}^{\ast}\cos^{2}\left(  \tilde{\alpha}\right)  -2\tilde
{A}\tilde{B}^{\ast}\sin^{2}\left(  \tilde{\alpha}\right) \nonumber\\
&  =\left(  \left\vert \tilde{A}\right\vert ^{2}+\left\vert \tilde
{B}\right\vert ^{2}-2\tilde{A}\tilde{B}^{\ast}\right)  \sin^{2}\left(
\tilde{\alpha}\right)  +\left(  \tilde{A}^{2}+\tilde{B}^{2}+2\tilde{A}%
\tilde{B}^{\ast}\right)  \cos^{2}\left(  \tilde{\alpha}\right) \nonumber\\
&  =\left\vert \tilde{A}-\tilde{B}\right\vert ^{2}\sin^{2}\left(
\tilde{\alpha}\right)  +\left\vert \tilde{A}+\tilde{B}\right\vert ^{2}\cos
^{2}\left(  \tilde{\alpha}\right)  \text{,}%
\end{align}
that is,%
\begin{equation}
\mathcal{P}_{\left\vert s\right\rangle \rightarrow\left\vert w\right\rangle
}\left(  \tilde{\alpha}\right)  =\left\vert \tilde{A}-\tilde{B}\right\vert
^{2}\sin^{2}\left(  \tilde{\alpha}\right)  +\left\vert \tilde{A}+\tilde
{B}\right\vert ^{2}\cos^{2}\left(  \tilde{\alpha}\right)  \text{.}
\label{fess}%
\end{equation}
Finally, using Eqs. (\ref{newroba}), (\ref{adef}), (\ref{bnew}), and
(\ref{anew}), $\mathcal{P}_{\left\vert s\right\rangle \rightarrow\left\vert
w\right\rangle }\left(  \tilde{\alpha}\right)  $ in Eq. (\ref{fess}) becomes%
\begin{equation}
\mathcal{P}_{\left\vert s\right\rangle \rightarrow\left\vert w\right\rangle
}\left(  t\right)  =x^{2}\cos^{2}\left(  \sqrt{\frac{h_{12}h_{21}}{\hslash
^{2}}+\frac{\left(  h_{11}-h_{22}\right)  ^{2}}{4\hslash^{2}}}t\right)
+\frac{\left\vert \frac{1}{2}\frac{h_{11}-h_{22}}{\hslash}x+\frac{h_{12}%
}{\hslash}\sqrt{1-x^{2}}\right\vert ^{2}}{\frac{h_{12}h_{21}}{\hslash^{2}%
}+\frac{\left(  h_{11}-h_{22}\right)  ^{2}}{4\hslash^{2}}}\sin^{2}\left(
\sqrt{\frac{h_{12}h_{21}}{\hslash^{2}}+\frac{\left(  h_{11}-h_{22}\right)
^{2}}{4\hslash^{2}}}t\right)  \text{.} \label{mn}%
\end{equation}
Appendix A ends with the derivation of the exact temporal behavior of\textbf{
}$\mathcal{P}_{\left\vert s\right\rangle \rightarrow\left\vert w\right\rangle
}\left(  t\right)  $ in Eq. (\ref{mn}).

\section{Time-dependent perturbation}

In this Appendix, we briefly review the concept of interaction representation
in the context of time-dependent perturbations. Furthermore, we comment on the
complexity of finding quantum mechanical probability amplitudes in an exact
analytical manner when the perturbation depends on time. The material
presented in this Appendix can be helpful to further clarify our work in
Section III.

Assume that a two-state quantum system is subject to a time-dependent
Hamiltonian evolution specified by $\mathcal{H}^{\left(  \text{full}\right)
}\left(  t\right)  $,%
\begin{equation}
\mathcal{H}^{\left(  \text{full}\right)  }\left(  t\right)  \overset
{\text{def}}{=}\mathcal{H}_{0}^{\left(  \text{free}\right)  }+\mathcal{V}%
^{\left(  \text{interaction}\right)  }\left(  t\right)  \text{,}%
\end{equation}
where $\mathcal{H}_{0}^{\left(  \text{free}\right)  }$ denotes the
time-independent free Hamiltonian while $\mathcal{V}^{\left(
\text{interaction}\right)  }\left(  t\right)  $ is the time-dependent
interaction potential. Furthermore, let $\mathcal{H}_{0}^{\left(
\text{free}\right)  }\left\vert E_{k}\right\rangle =E_{k}\left\vert
E_{k}\right\rangle $ for any $1\leq k\leq2$. The energy eigenstates $\left\{
\left\vert E_{k}\right\rangle \right\}  $ are such that $\left\langle
E_{n}|E_{m}\right\rangle =\delta_{nm}$ for any $1\leq n$, $m\leq2$. Then, the
temporal evolution of a quantum state $\left\vert s\right\rangle
\overset{\text{def}}{=}\left\vert s\left(  0\right)  \right\rangle $,%
\begin{equation}
\left\vert s\left(  0\right)  \right\rangle \overset{\text{def}}{=}%
{\displaystyle\sum\limits_{k=1}^{2}}
c_{k}\left(  0\right)  \left\vert E_{k}\right\rangle \text{,}%
\end{equation}
can be described as,%
\begin{equation}
\left\vert s\left(  t\right)  \right\rangle \overset{\text{def}}{=}%
{\displaystyle\sum\limits_{k=1}^{2}}
c_{k}\left(  t\right)  e^{-\frac{i}{\hslash}E_{k}t}\left\vert E_{k}%
\right\rangle \text{.} \label{st}%
\end{equation}
From Eq. (\ref{st}), observe that the presence of a nonzero interaction
potential implies that the coefficients $c_{k}\left(  t\right)  $ are
time-dependent. Instead, if $\mathcal{V}^{\left(  \text{interaction}\right)
}\left(  t\right)  =0$ then $c_{k}\left(  t\right)  =c_{k}\left(  0\right)  $.
To find an equation satisfied by the coefficients $c_{k}\left(  t\right)  $,
we proceed as follows \cite{sakurai, picasso}. The interaction representation
of a quantum state that at time $t=0$ is given by $\left\vert s\left(
0\right)  \right\rangle $ is defined as,%
\begin{equation}
\left\vert s\left(  t\right)  \right\rangle _{\text{I}}\overset{\text{def}}%
{=}e^{\frac{i}{\hslash}\mathcal{H}_{0}^{\left(  \text{free}\right)  }%
t}\left\vert s\left(  0\right)  \right\rangle _{\text{S}}\text{.} \label{IR}%
\end{equation}
The symbols \textquotedblleft I\textquotedblright\ and \textquotedblleft
S\textquotedblright\ in\ Eq. (\ref{IR}) denote the interaction and the
Schr\"{o}dinger representations, respectively. Note that $\left\vert s\left(
0\right)  \right\rangle _{\text{S}}=\left\vert s\left(  0\right)
\right\rangle _{\text{I}}=\left\vert s\left(  0\right)  \right\rangle $.
Multiplying both sides of Eq. (\ref{IR}) by $i\hslash$, differentiating with
respect to $t$, and recalling that $i\hslash\partial_{t}\left\vert s\left(
0\right)  \right\rangle _{\text{S}}=\mathcal{H}_{0}^{\left(  \text{free}%
\right)  }\left\vert s\left(  0\right)  \right\rangle _{\text{S}}$, we obtain%
\begin{equation}
i\hslash\partial_{t}\left\vert s\left(  t\right)  \right\rangle _{\text{I}%
}=\mathcal{V}_{\text{I}}\left\vert s\left(  t\right)  \right\rangle
_{\text{I}} \label{VI}%
\end{equation}
The operator $\mathcal{V}_{\text{I}}$ in Eq. (\ref{VI}) is the interaction
representation of the interaction potential $\mathcal{V}^{\left(
\text{interaction}\right)  }$ and is defined as,%
\begin{equation}
\mathcal{V}_{\text{I}}\overset{\text{def}}{=}e^{\frac{i}{\hslash}%
\mathcal{H}_{0}^{\left(  \text{free}\right)  }t}\mathcal{V}^{\left(
\text{interaction}\right)  }e^{-\frac{i}{\hslash}\mathcal{H}_{0}^{\left(
\text{free}\right)  }t}\text{.}%
\end{equation}
Multiplying both sides of Eq. (\ref{VI}) for $\left\langle E_{k}\right\vert $
from the left and using the completeness relation,%
\begin{equation}%
{\displaystyle\sum\limits_{n=1}^{2}}
\left\vert E_{n}\right\rangle \left\langle E_{n}\right\vert =\mathcal{I}%
\text{,}%
\end{equation}
with $\mathcal{I}$ denoting the identity operator, we obtain%
\begin{equation}
i\hslash\partial_{t}\left\langle E_{k}|s\left(  t\right)  \right\rangle
_{\text{I}}=%
{\displaystyle\sum\limits_{n=1}^{2}}
\left\langle E_{k}|\mathcal{V}_{\text{I}}|E_{n}\right\rangle \left\langle
E_{n}|s\left(  t\right)  \right\rangle _{\text{I}}\text{.} \label{quisopra1}%
\end{equation}
Observe that $c_{k}\left(  t\right)  \overset{\text{def}}{=}\left\langle
E_{k}|s\left(  t\right)  \right\rangle _{\text{I}}$, and%
\begin{equation}
\left\langle E_{k}|\mathcal{V}_{\text{I}}|E_{n}\right\rangle =\left\langle
E_{k}|\mathcal{V}^{\left(  \text{interaction}\right)  }\left(  t\right)
|E_{n}\right\rangle e^{\frac{i}{\hslash}\left(  E_{k}-E_{n}\right)
t}=\mathcal{V}_{kn}\left(  t\right)  e^{i\omega_{kn}t}\text{,}
\label{quisopra}%
\end{equation}
where $\mathcal{V}_{kn}\left(  t\right)  \overset{\text{def}}{=}\left\langle
E_{k}|\mathcal{V}^{\left(  \text{interaction}\right)  }\left(  t\right)
|E_{n}\right\rangle $ and $\omega_{kn}\overset{\text{def}}{=}\left(
E_{k}-E_{n}\right)  /\hslash$. Finally, using Eq. (\ref{quisopra}), Eq.
(\ref{quisopra1}) becomes%
\begin{equation}
i\hslash\partial_{t}c_{k}\left(  t\right)  =%
{\displaystyle\sum\limits_{n=1}^{2}}
\mathcal{V}_{kn}\left(  t\right)  e^{i\omega_{kn}t}c_{n}\left(  t\right)
\text{.} \label{odesystem}%
\end{equation}
Equation (\ref{odesystem}) is a system of coupled first order ordinary
differential equations with time-dependent coefficients that need to be
integrated in order to find the coefficients $c_{k}\left(  t\right)  $. We
remark that once the time-dependent quantum mechanical probability amplitudes
$\left\{  c_{k}\left(  t\right)  \right\}  $ are obtained, the transition
probabilities $\mathcal{P}_{\left\vert s\right\rangle \rightarrow\left\vert
E_{1}\right\rangle }\left(  t\right)  $ and $\mathcal{P}_{\left\vert
s\right\rangle \rightarrow\left\vert E_{1}\right\rangle }\left(  t\right)  $
are given by%
\begin{equation}
\mathcal{P}_{\left\vert s\right\rangle \rightarrow\left\vert E_{1}%
\right\rangle }\left(  t\right)  \overset{\text{def}}{=}\left[  c_{1}\left(
t\right)  \right]  \left[  c_{1}\left(  t\right)  \right]  ^{\ast}=\left\vert
c_{1}\left(  t\right)  \right\vert ^{2}\text{, and }\mathcal{P}_{\left\vert
s\right\rangle \rightarrow\left\vert E_{2}\right\rangle }\left(  t\right)
\overset{\text{def}}{=}\left[  c_{2}\left(  t\right)  \right]  \left[
c_{2}\left(  t\right)  \right]  ^{\ast}=\left\vert c_{2}\left(  t\right)
\right\vert ^{2}\text{,} \label{recallyou}%
\end{equation}
respectively.

In the remainder of this Appendix, we provide further technical details on the
structure of the ODEs that arise from Eq. (\ref{odesystem}). Recall that the
probability amplitudes $c_{1}\left(  t\right)  $ and $c_{2}\left(  t\right)  $
satisfy the following differential equations,%
\begin{equation}
i\hslash\dot{c}_{1}=\mathcal{V}_{11}c_{1}+\mathcal{V}_{12}e^{i\omega_{12}%
t}c_{2}\text{,} \label{de1}%
\end{equation}
and,%
\begin{equation}
i\hslash\dot{c}_{2}=\mathcal{V}_{21}e^{i\omega_{21}t}c_{1}+\mathcal{V}%
_{22}c_{2}\text{,} \label{de2}%
\end{equation}
respectively, where $\omega_{12}=-\omega_{21}$ since $\omega_{21}%
\overset{\text{def}}{=}(E_{2}-E_{1})/\hslash$ and $\dot{c}_{i}\overset
{\text{def}}{=}dc_{i}/dt$ with $i\in\left\{  1\text{, }2\right\}  $.
Differentiating both sides of Eq. (\ref{de1}) with respect to time, we obtain%
\begin{equation}
i\hslash\ddot{c}_{1}-\mathcal{\dot{V}}_{11}c_{1}-\mathcal{V}_{11}\dot{c}%
_{1}=\left(  \mathcal{\dot{V}}_{12}+i\omega_{12}\mathcal{V}_{12}\right)
e^{i\omega_{12}t}c_{2}+\mathcal{V}_{12}e^{i\omega_{12}t}\dot{c}_{2}
\label{de3}%
\end{equation}
From Eq. (\ref{de1}), it is found that $c_{2}$ equals%
\begin{equation}
c_{2}=\frac{i\hslash}{\mathcal{V}_{12}}e^{-i\omega_{12}t}\dot{c}_{1}%
-\frac{\mathcal{V}_{11}}{\mathcal{V}_{12}}e^{-i\omega_{12}t}c_{1}\text{.}
\label{de4}%
\end{equation}
Using Eqs. (\ref{de2}) and (\ref{de4}), $\dot{c}_{2}$ can be rewritten as%
\begin{equation}
\dot{c}_{2}=\frac{\mathcal{V}_{22}}{\mathcal{V}_{12}}e^{-i\omega_{12}t}\dot
{c}_{1}-\frac{\mathcal{V}_{11}\mathcal{V}_{22}}{i\hslash\mathcal{V}_{12}%
}e^{-i\omega_{12}t}c_{1}+\frac{\mathcal{V}_{21}}{i\hslash}e^{-i\omega_{12}%
t}c_{1}\text{.} \label{de5}%
\end{equation}
Finally, substituting Eqs. (\ref{de4}) and (\ref{de5}) into Eq. (\ref{de3}),
we arrive at%
\begin{equation}
\ddot{c}_{1}-\left(  \frac{\mathcal{\dot{V}}_{12}+i\omega_{12}\mathcal{V}%
_{12}}{\mathcal{V}_{12}}+\frac{\mathcal{V}_{11}+\mathcal{V}_{22}}{i\hslash
}\right)  \dot{c}_{1}+\left[  \frac{\mathcal{V}_{11}\left(  \mathcal{\dot{V}%
}_{12}+i\omega_{12}\mathcal{V}_{12}\right)  }{i\hslash\mathcal{V}_{12}}%
-\frac{\mathcal{\dot{V}}_{11}}{i\hslash}-\frac{\left(  \mathcal{V}%
_{11}\mathcal{V}_{22}-\mathcal{V}_{21}\mathcal{V}_{12}\right)  }{\hslash^{2}%
}\right]  c_{1}=0\text{.} \label{de6}%
\end{equation}
Proceeding along this same line of reasoning, we readily obtain%
\begin{equation}
\ddot{c}_{2}-\left(  \frac{\mathcal{\dot{V}}_{21}+i\omega_{21}\mathcal{V}%
_{21}}{\mathcal{V}_{21}}+\frac{\mathcal{V}_{11}+\mathcal{V}_{22}}{i\hslash
}\right)  \dot{c}_{2}+\left[  \frac{\mathcal{V}_{22}\left(  \mathcal{\dot{V}%
}_{21}+i\omega_{21}\mathcal{V}_{21}\right)  }{i\hslash\mathcal{V}_{21}}%
-\frac{\mathcal{\dot{V}}_{22}}{i\hslash}-\frac{\left(  \mathcal{V}%
_{11}\mathcal{V}_{22}-\mathcal{V}_{21}\mathcal{V}_{12}\right)  }{\hslash^{2}%
}\right]  c_{2}=0\text{.} \label{de7}%
\end{equation}
In general, Eqs. (\ref{de6})\ and (\ref{de7}) are second-order linear ordinary
differential equations with non-constant coefficients. Observe that in the
special case in which $\mathcal{V}_{11}=0$, $\mathcal{V}_{22}=0$,
$\mathcal{V}_{12}=\Gamma e^{i\omega t}$, and $\mathcal{V}_{21}=\Gamma
e^{-i\omega t}$, Eqs. (\ref{de6}) and (\ref{de7}) become%
\begin{equation}
\ddot{c}_{1}-i\left(  \omega-\omega_{21}\right)  \dot{c}_{1}+\frac{\Gamma^{2}%
}{\hslash^{2}}c_{1}=0\text{,} \label{de6a}%
\end{equation}
and,%
\begin{equation}
\ddot{c}_{2}+i\left(  \omega-\omega_{21}\right)  \dot{c}_{2}+\frac{\Gamma^{2}%
}{\hslash^{2}}c_{2}=0\text{,} \label{de7a}%
\end{equation}
respectively. Unlike Eqs. (\ref{de6}) and (\ref{de7}), Eqs. (\ref{de6a}) and
(\ref{de7a}) are second-order linear ordinary differential equations with
constant coefficients. While this set of two ODEs will be explicitly
integrated in Appendix C, the exact analytical integration of Eqs.
(\ref{de6})\ and (\ref{de7}) can be generally quite challenging.

\section{Integration of Eq. (\ref{system1})}

In this Appendix, we integrate the system of coupled ODEs in Eq.
(\ref{system1}) that appear in Section III.

After some algebraic manipulations, the two relations in Eq. (\ref{system1}%
)\ lead to the following second order ordinary differential equation with
time-independent coefficients for $c_{2}$,%
\begin{equation}
\ddot{c}_{2}+i\left(  \omega-\omega_{21}\right)  \dot{c}_{2}+\frac{\Gamma^{2}%
}{\hslash^{2}}c_{2}=0\text{.} \label{c2}%
\end{equation}
We note that a particular solution of Eq. (\ref{c2}) can be written as
$c_{2}\left(  t\right)  =Ae^{\alpha t}$ with $A$ and $\alpha$ denoting two
$\emph{complex}$ constants. In particular, the constant $\alpha$ can be found
by substituting the condition $c_{2}\left(  t\right)  =Ae^{\alpha t}$ into Eq.
(\ref{c2}). We obtain the following constraint equation,%
\begin{equation}
\alpha^{2}+\alpha i\left(  \omega-\omega_{21}\right)  +\frac{\Gamma^{2}%
}{\hslash^{2}}=0\text{,}%
\end{equation}
which, in turn, leads to the two \emph{complex} roots%
\begin{equation}
\alpha_{\pm}\overset{\text{def}}{=}-\frac{i}{2}\left(  \omega-\omega
_{21}\right)  \pm i\sqrt{\frac{\Gamma^{2}}{\hslash^{2}}+\frac{\left(
\omega-\omega_{21}\right)  ^{2}}{4}}\text{.} \label{alpha}%
\end{equation}
Therefore, the general solution of Eq. (\ref{c2}) can be formally written as%
\begin{equation}
c_{2}\left(  t\right)  =\mathcal{A}e^{\alpha_{+}t}+\mathcal{B}e^{\alpha_{-}%
t}\text{,} \label{c2g}%
\end{equation}
where the constants $\mathcal{A}$ and $\mathcal{B}$ belong to $%
\mathbb{C}
$. Substituting Eq. (\ref{alpha}) into Eq. (\ref{c2g}), $c_{2}\left(
t\right)  $ becomes%
\begin{equation}
c_{2}\left(  t\right)  =e^{-\frac{i}{2}\left(  \omega-\omega_{21}\right)
t}\left[  \mathcal{A}\exp\left(  i\sqrt{\frac{\Gamma^{2}}{\hslash^{2}}%
+\frac{\left(  \omega-\omega_{21}\right)  ^{2}}{4}}t\right)  +\mathcal{B}%
\exp\left(  -i\sqrt{\frac{\Gamma^{2}}{\hslash^{2}}+\frac{\left(  \omega
-\omega_{21}\right)  ^{2}}{4}}t\right)  \right]  \text{,} \label{c2n}%
\end{equation}
where $\exp\left(  \cdot\right)  $ denotes the exponential function. Recalling
that $e^{ix}\overset{\text{def}}{=}\cos\left(  x\right)  +i\sin\left(
x\right)  $ for any $x$ in $%
\mathbb{R}
$, after some algebraic manipulations, we deduce that $c_{2}\left(  t\right)
$ in Eq. (\ref{c2n}) can be recast as%
\begin{equation}
c_{2}\left(  t\right)  =e^{-\frac{i}{2}\left(  \omega-\omega_{21}\right)
t}\left[  \mathcal{C}\cos\left(  \sqrt{\frac{\Gamma^{2}}{\hslash^{2}}%
+\frac{\left(  \omega-\omega_{21}\right)  ^{2}}{4}}t\right)  +\mathcal{D}%
\sin\left(  \sqrt{\frac{\Gamma^{2}}{\hslash^{2}}+\frac{\left(  \omega
-\omega_{21}\right)  ^{2}}{4}}t\right)  \right]  \text{,}%
\end{equation}
with $\mathcal{C}\overset{\text{def}}{=}\mathcal{A}+\mathcal{B}$ and
$\mathcal{D}\overset{\text{def}}{=}i(\mathcal{A}-\mathcal{B})$. Imposing that
$c_{2}\left(  0\right)  =x$, we find that $\mathcal{C}=x$ and, as a
consequence,%
\begin{equation}
c_{2}\left(  t\right)  =e^{-\frac{i}{2}\left(  \omega-\omega_{21}\right)
t}\left[  x\cos\left(  \sqrt{\frac{\Gamma^{2}}{\hslash^{2}}+\frac{\left(
\omega-\omega_{21}\right)  ^{2}}{4}}t\right)  +\mathcal{D}\sin\left(
\sqrt{\frac{\Gamma^{2}}{\hslash^{2}}+\frac{\left(  \omega-\omega_{21}\right)
^{2}}{4}}t\right)  \right]  \text{.} \label{c2final}%
\end{equation}
For the sake of notational simplicity, let us introduce the following
\emph{real} quantity%
\begin{equation}
\Omega\overset{\text{def}}{=}\sqrt{\frac{\Gamma^{2}}{\hslash^{2}}%
+\frac{\left(  \omega-\omega_{21}\right)  ^{2}}{4}}\text{.} \label{put}%
\end{equation}
Combining the first relation in Eq. (\ref{system1}) with Eqs. (\ref{c2final})
and (\ref{put}), it can be verified that $c_{1}\left(  t\right)  $ satisfies
the condition%
\begin{equation}
c_{1}\left(  t\right)  =-i\frac{\Gamma}{\hslash}\int e^{\frac{i}{2}\left(
\omega-\omega_{21}\right)  t}\left[  x\cos\left(  \Omega t\right)
+\mathcal{D}\sin\left(  \Omega t\right)  \right]  dt\text{.} \label{c1}%
\end{equation}
Observe that for any \emph{complex} coefficients $a$ and $b$, up to an
unimportant constant of integration, we have%
\begin{equation}
\int e^{at}\sin\left(  bt\right)  dt=\frac{e^{at}}{a^{2}+b^{2}}\left[
a\sin\left(  bt\right)  -b\cos\left(  bt\right)  \right]  \text{,}
\label{chist1}%
\end{equation}
and,%
\begin{equation}
\int e^{at}\cos\left(  bt\right)  dt=\frac{e^{at}}{a^{2}+b^{2}}\left[
b\sin\left(  bt\right)  +a\cos\left(  bt\right)  \right]  \text{.}
\label{chist2}%
\end{equation}
Using Eqs. (\ref{chist1}) and (\ref{chist2}), integration of Eq. (\ref{c1})
yields%
\begin{equation}
c_{1}\left(  t\right)  =-i\frac{\hslash}{\Gamma}e^{i\left(  \omega-\omega
_{21}\right)  t}\left\{  \left[  \Omega x+\frac{i}{2}\left(  \omega
-\omega_{21}\right)  \mathcal{D}\right]  \sin\left(  \Omega t\right)  -\left[
\Omega\mathcal{D-}\frac{i}{2}\left(  \omega-\omega_{21}\right)  x\right]
\cos\left(  \Omega t\right)  \right\}  \text{.} \label{c1final}%
\end{equation}
To find the integration constant $\mathcal{D}$, we impose that $c_{1}\left(
0\right)  =\sqrt{1-x^{2}}$. We therefore\textbf{ }obtain,%
\begin{equation}
\mathcal{D}\overset{\text{def}}{=}i\frac{\frac{\left(  \omega-\omega
_{21}\right)  }{2}x-\frac{\Gamma}{\hslash}\sqrt{1-x^{2}}}{\sqrt{\frac
{\Gamma^{2}}{\hslash^{2}}+\frac{\left(  \omega-\omega_{21}\right)  ^{2}}{4}}%
}\text{.} \label{Deq}%
\end{equation}
At this point, we recall that the transition probabilities $\mathcal{P}%
_{\left\vert s\right\rangle \rightarrow\left\vert E_{1}\right\rangle }\left(
t\right)  $ and $\mathcal{P}_{\left\vert s\right\rangle \rightarrow\left\vert
E_{2}\right\rangle }\left(  t\right)  $ are defined as $\mathcal{P}%
_{\left\vert s\right\rangle \rightarrow\left\vert E_{1}\right\rangle }\left(
t\right)  \overset{\text{def}}{=}\left\vert c_{1}\left(  t\right)  \right\vert
^{2}$ and $\mathcal{P}_{\left\vert s\right\rangle \rightarrow\left\vert
E_{2}\right\rangle }\left(  t\right)  \overset{\text{def}}{=}\left\vert
c_{2}\left(  t\right)  \right\vert ^{2}$, respectively. The quantum mechanical
probability amplitudes $c_{1}\left(  t\right)  \overset{\text{def}}%
{=}\left\langle E_{1}|s\left(  t\right)  \right\rangle _{\text{I}}$ and
$c_{2}\left(  t\right)  \overset{\text{def}}{=}\left\langle E_{2}|s\left(
t\right)  \right\rangle _{\text{I}}$ are given in Eqs. (\ref{c1final}) and
(\ref{c2final}), respectively. Furthermore, the quantities $\Omega$ and
$\mathcal{D}$ are defined in Eqs. (\ref{put}) and (\ref{Deq}), respectively.
After some algebra, we find that the transition probabilities $\mathcal{P}%
_{\left\vert s\right\rangle \rightarrow\left\vert E_{1}\right\rangle }\left(
t\right)  $ and $\mathcal{P}_{\left\vert s\right\rangle \rightarrow\left\vert
E_{2}\right\rangle }\left(  t\right)  $ are given by%
\begin{equation}
\mathcal{P}_{\left\vert s\right\rangle \rightarrow\left\vert E_{1}%
\right\rangle }\left(  t\right)  =\left(  1-x^{2}\right)  \cos^{2}\left(
\sqrt{\frac{\Gamma^{2}}{\hslash^{2}}+\frac{\left(  \omega-\omega_{21}\right)
^{2}}{4}}t\right)  +\left\{  1-\left[  \frac{\frac{\left(  \omega-\omega
_{21}\right)  }{2}x-\frac{\Gamma}{\hslash}\sqrt{1-x^{2}}}{\sqrt{\frac
{\Gamma^{2}}{\hslash^{2}}+\frac{\left(  \omega-\omega_{21}\right)  ^{2}}{4}}%
}\right]  ^{2}\right\}  \sin^{2}\left(  \sqrt{\frac{\Gamma^{2}}{\hslash^{2}%
}+\frac{\left(  \omega-\omega_{21}\right)  ^{2}}{4}}t\right)  \text{,}
\label{pio1}%
\end{equation}
and,%
\begin{equation}
\mathcal{P}_{\left\vert s\right\rangle \rightarrow\left\vert E_{2}%
\right\rangle }\left(  t\right)  =x^{2}\cos^{2}\left(  \sqrt{\frac{\Gamma^{2}%
}{\hslash^{2}}+\frac{\left(  \omega-\omega_{21}\right)  ^{2}}{4}}t\right)
+\left[  \frac{\frac{\left(  \omega-\omega_{21}\right)  }{2}x-\frac{\Gamma
}{\hslash}\sqrt{1-x^{2}}}{\sqrt{\frac{\Gamma^{2}}{\hslash^{2}}+\frac{\left(
\omega-\omega_{21}\right)  ^{2}}{4}}}\right]  ^{2}\sin^{2}\left(  \sqrt
{\frac{\Gamma^{2}}{\hslash^{2}}+\frac{\left(  \omega-\omega_{21}\right)  ^{2}%
}{4}}t\right)  \text{,} \label{pio2}%
\end{equation}
respectively. Appendix C ends with the derivations of the exact temporal
behaviors of\textbf{ }$\mathcal{P}_{\left\vert s\right\rangle \rightarrow
\left\vert E_{1}\right\rangle }\left(  t\right)  $ and $\mathcal{P}%
_{\left\vert s\right\rangle \rightarrow\left\vert E_{2}\right\rangle }\left(
t\right)  $ in Eqs. (\ref{pio1}) and (\ref{pio2}), respectively.

\section{Probability amplitudes in the Zener case}

In this Appendix, we present some technical details on the mathematical scheme
employed by Zener in Ref. \cite{zener32a} and used by Perez and Romanelli in
Ref. \cite{romanelli07} as mentioned in Section IV. This approach helps
with\textbf{ }solving a special case of Eqs. (\ref{de6}) and (\ref{de7}). In
particular, Zener considered a special case of Eq. (\ref{de6}) given by%
\begin{equation}
\ddot{c}_{1}+i\alpha t\dot{c}_{1}+f^{2}c_{1}=0\text{,} \label{de8}%
\end{equation}
with $\alpha$ and $f^{2}$ denoting two \emph{real} constants. Upon a suitable
change of the dependent variable $c_{1}$,%
\begin{equation}
c_{1}\left(  t\right)  \rightarrow c_{1}\left(  t\right)  \overset{\text{def}%
}{=}e^{-i\frac{\alpha}{4}t^{2}}u_{1}\left(  t\right)  \text{,}%
\end{equation}
Eq. (\ref{de8}) becomes%
\begin{equation}
\ddot{u}_{1}+\left(  f^{2}-i\frac{\alpha}{2}+\frac{\alpha^{2}}{4}t^{2}\right)
u_{1}=0\text{.} \label{de9}%
\end{equation}
Multiplying Eq. (\ref{de9}) by $i/\alpha$ and defining,%
\begin{equation}
n\overset{\text{def}}{=}\frac{if^{2}}{\alpha}\text{, and }z\overset
{\text{def}}{=}\sqrt{\alpha}e^{-i\frac{\pi}{4}}t\text{,}%
\end{equation}
Eq. (\ref{de9}) can be recast as%
\begin{equation}
\frac{i}{\alpha}\ddot{u}_{1}\left(  t\right)  +\left(  n+\frac{1}{2}-\frac
{1}{4}z^{2}\right)  u_{1}\left(  t\right)  =0\text{.} \label{de10}%
\end{equation}
In terms of the independent variable $z$, Eq. (\ref{de10}) becomes%
\begin{equation}
u_{1}^{\prime\prime}\left(  z\right)  +\left(  n+\frac{1}{2}-\frac{1}{4}%
z^{2}\right)  u_{1}\left(  z\right)  =0\text{,} \label{de11}%
\end{equation}
where $u_{1}^{\prime}\overset{\text{def}}{=}du_{1}/dz$. Finally, we recognize
that\textbf{ }Eq. (\ref{de11})\ is a Weber differential equation whose general
solution $u_{1}\left(  z\text{; }n\right)  $ can be written as a superposition
of independent solutions $\mathcal{D}_{n}\left(  z\right)  $ and
$\mathcal{D}_{-n-1}\left(  iz\right)  $ as follows%
\begin{equation}
u_{1}\left(  z\text{; }n\right)  =\mathrm{a}\mathcal{D}_{n}\left(  z\right)
+\mathrm{b}\mathcal{D}_{-n-1}\left(  iz\right)  \text{,} \label{de12}%
\end{equation}
where $\mathrm{a}$ and $\mathrm{b}$ are integration constants. The functions
$\mathcal{D}_{n}\left(  z\right)  $ in Eq. (\ref{de12}) are known as parabolic
cylinder functions \cite{irene} and are defined as,%
\begin{equation}
\mathcal{D}_{n}\left(  z\right)  \overset{\text{def}}{=}\frac{2^{\frac{n}%
{2}+\frac{1}{4}}}{\sqrt{z}}\mathcal{W}_{\frac{n}{2}+\frac{1}{4}\text{, }%
-\frac{1}{4}}\left(  \frac{z^{2}}{2}\right)  \text{.} \label{de13}%
\end{equation}
The functions $\mathcal{W}_{k\text{, }m}\left(  z\right)  $ in Eq.
(\ref{de13}) are known as the Whittaker functions \cite{irene} and are defined
as,%
\begin{equation}
\mathcal{W}_{k\text{, }m}\left(  z\right)  \overset{\text{def}}{=}%
z^{m+\frac{1}{2}}e^{-\frac{z}{2}}%
{\displaystyle\sum\limits_{n=0}^{\infty}}
\frac{\left(  m-k+\frac{1}{2}\right)  _{n}}{n!\left(  2m+1\right)  _{n}}%
z^{n}\text{,}%
\end{equation}
where $\left(  z\right)  _{n}$ denotes the Pochhammer symbol defined as
\cite{irene},%
\begin{equation}
\left(  z\right)  _{n}\overset{\text{def}}{=}\frac{\Gamma\left(  z+n\right)
}{\Gamma\left(  z\right)  }=\frac{\left(  z+n-1\right)  !}{\left(  z-1\right)
!}=z\left(  z+1\right)  \text{...}\left(  z+n-1\right)  \text{.} \label{zetan}%
\end{equation}
The quantity $\Gamma$ in Eq. (\ref{zetan}) is the Euler gamma function. For
further details on the Weber equation, the parabolic cylinder functions, and
the Whittaker functions, we refer to Ref. \cite{irene}.

\section{Analytically solvable quantum two-level systems}

In this Appendix, we present a few preliminary mathematical remarks inspired
by the recent investigations presented by Messina and collaborators in Refs.
\cite{messina14,grimaudo18}. These remarks can be helpful to further clarify
our work presented in Section V.

The matrix representation of the unitary evolution operator $\mathcal{U}%
\left(  t\right)  $ generated by the time-dependent Hamiltonian $\mathcal{H}%
\left(  t\right)  $ in Eq. (\ref{angel}) with $i\hslash\mathcal{\dot{U}%
}\left(  t\right)  =\mathcal{H}\left(  t\right)  \mathcal{U}\left(  t\right)
$ where $\mathcal{\dot{U}}\overset{\text{def}}{=}\partial_{t}\mathcal{U}$ can
be described as,%
\begin{equation}
\left[  \mathcal{U}\left(  t\right)  \right]  _{\mathcal{B}_{\text{canonical}%
}}\overset{\text{def}}{=}\left(
\begin{array}
[c]{cc}%
\alpha\left(  t\right)   & \beta\left(  t\right)  \\
-\beta^{\ast}\left(  t\right)   & \alpha^{\ast}\left(  t\right)
\end{array}
\right)  \text{.}\label{evolutiono}%
\end{equation}
Note that unitarity requires that the \emph{complex} probability amplitudes
$\alpha\left(  t\right)  $ and $\beta\left(  t\right)  $ satisfy the
normalization condition,%
\begin{equation}
\left\vert \alpha\left(  t\right)  \right\vert ^{2}+\left\vert \beta\left(
t\right)  \right\vert ^{2}=1\text{.}\label{unitarity}%
\end{equation}
Observe that the temporal evolution of a quantum source state $\left\vert
s\right\rangle $,%
\begin{equation}
\left\vert s\right\rangle \overset{\text{def}}{=}x\left\vert w\right\rangle
+\sqrt{1-x^{2}}\left\vert r\right\rangle \text{,}%
\end{equation}
under the unitary evolution operator $\mathcal{U}\left(  t\right)  $ in Eq.
(\ref{evolutiono}) can be described in terms of the following mapping,%
\begin{equation}
\binom{x}{\sqrt{1-x^{2}}}\rightarrow\binom{\alpha\left(  t\right)
x+\beta\left(  t\right)  \sqrt{1-x^{2}}}{-\beta^{\ast}\left(  t\right)
x+\alpha^{\ast}\left(  t\right)  \sqrt{1-x^{2}}}\text{.}%
\end{equation}
Therefore, the probability that the source state $\left\vert s\right\rangle $
transitions into the target state $\left\vert w\right\rangle $ under
$\mathcal{U}\left(  t\right)  $ is given by,%
\begin{equation}
\mathcal{P}_{\left\vert s\right\rangle \rightarrow\left\vert w\right\rangle
}\left(  t\right)  \overset{\text{def}}{=}\left\vert \left\langle
w|\mathcal{U}\left(  t\right)  |s\right\rangle \right\vert ^{2}\text{,}%
\end{equation}
that is,%
\begin{equation}
\mathcal{P}_{\left\vert s\right\rangle \rightarrow\left\vert w\right\rangle
}\left(  t\right)  =\left\vert \alpha\left(  t\right)  \right\vert ^{2}%
x^{2}+\left\vert \beta\left(  t\right)  \right\vert ^{2}\left(  1-x^{2}%
\right)  +\left[  \alpha\left(  t\right)  \beta^{\ast}\left(  t\right)
+\alpha^{\ast}\left(  t\right)  \beta\left(  t\right)  \right]  x\sqrt
{1-x^{2}}\text{,}\label{TP1}%
\end{equation}
with $\alpha\left(  t\right)  \beta^{\ast}\left(  t\right)  +\alpha^{\ast
}\left(  t\right)  \beta\left(  t\right)  =2\left[  \alpha_{\text{R}}\left(
t\right)  \beta_{\text{R}}\left(  t\right)  +\alpha_{\text{I}}\left(
t\right)  \beta_{\text{I}}\left(  t\right)  \right]  $ being a \emph{real}
quantity. In the last statement, we assumed that $\alpha\left(  t\right)  $
and $\beta\left(  t\right)  $ could be rewritten as,
\begin{equation}
\alpha\left(  t\right)  \overset{\text{def}}{=}\operatorname{Re}\left[
\alpha\left(  t\right)  \right]  +i\operatorname{Im}\left[  \alpha\left(
t\right)  \right]  =\alpha_{\text{R}}\left(  t\right)  +i\alpha_{\text{I}%
}\left(  t\right)  \text{,}%
\end{equation}
and,
\begin{equation}
\beta\left(  t\right)  \overset{\text{def}}{=}\operatorname{Re}\left[
\beta\left(  t\right)  \right]  +i\operatorname{Im}\left[  \beta\left(
t\right)  \right]  =\beta_{\text{R}}\left(  t\right)  +i\beta_{\text{I}%
}\left(  t\right)  \text{,}%
\end{equation}
respectively. From Eq. (\ref{TP1}), it is evident that in order\textbf{ }to
compute transition probabilities, one needs to know the evolution operator
$\mathcal{U}\left(  t\right)  $ in terms of the complex probability amplitudes
$\alpha\left(  t\right)  $ and $\beta\left(  t\right)  $. Ideally, having
fixed the fields $\omega\left(  t\right)  $ and $\Omega\left(  t\right)  $ in
Eq. (\ref{ham1}) due to the motivation of physical arguments, one would try to
integrate the coupled system of first order ordinary differential equations
with time-dependent coefficients generated by the relation $i\hslash
\mathcal{\dot{U}}\left(  t\right)  =\mathcal{H}\left(  t\right)
\mathcal{U}\left(  t\right)  $ with $\mathcal{U}\left(  0\right)
=\mathcal{I}$,%
\begin{equation}
i\hslash\dot{\alpha}\left(  t\right)  =\Omega\left(  t\right)  \alpha\left(
t\right)  -\omega\left(  t\right)  \beta^{\ast}\left(  t\right)  \text{, and
}i\hslash\dot{\beta}\left(  t\right)  =\omega\left(  t\right)  \alpha^{\ast
}\left(  t\right)  +\Omega\left(  t\right)  \beta\left(  t\right)
\text{,}\label{lodes}%
\end{equation}
where $\alpha\left(  0\right)  =1$ and $\beta\left(  0\right)  =0$.
Unfortunately, this general approach can rarely be solved in an exact manner.
If one is willing to specify only the functional form of one of the two fields
$\omega\left(  t\right)  $ and $\Omega\left(  t\right)  $ however, it is
possible to introduce clever mathematical schemes that allow to solve
analytically the system of ODEs in Eq. (\ref{lodes}), at least for certain
relevant functional forms of the non-fixed field \cite{messina14,grimaudo18}.
In what follows, we build a\textbf{ }great part of our discussion on the very
interesting works presented in Refs. \cite{messina14,grimaudo18}.

For the sake of reasoning, let us assume that we fix the field $\Omega\left(
t\right)  $. After some algebra, it can be shown that the second relation in
Eq. ((\ref{lodes})\ is solved by $\beta\left(  t\right)  $,%
\begin{equation}
\beta\left(  t\right)  =\frac{1}{i\hslash}\alpha\left(  t\right)  X\left(
t\right)  \text{,} \label{beta}%
\end{equation}
where $X\left(  t\right)  $ with $X\left(  0\right)  =0$ is a \emph{complex}
auxiliary function defined as,%
\begin{equation}
X\left(  t\right)  \overset{\text{def}}{=}\int_{0}^{t}\frac{\omega\left(
t^{\prime}\right)  }{\alpha^{2}\left(  t^{\prime}\right)  }dt^{\prime}\text{.}
\label{xxt}%
\end{equation}
From Eq. (\ref{xxt}), it follows that%
\begin{equation}
\omega\left(  t\right)  =\alpha^{2}\left(  t\right)  \dot{X}\left(  t\right)
\text{.} \label{omegaf}%
\end{equation}
Using Eqs. (\ref{unitarity}), (\ref{beta}), and (\ref{xxt}), it can be shown
that the first relation in Eq. (\ref{lodes}) leads to the following
differential equation for $\alpha\left(  t\right)  $,%
\begin{equation}
\dot{\alpha}\left(  t\right)  =-\left[  \frac{i}{\hslash}\Omega\left(
t\right)  +\frac{\dot{X}\left(  t\right)  X^{\ast}\left(  t\right)  }%
{\hslash^{2}+\left\vert X\left(  t\right)  \right\vert ^{2}}\right]
\alpha\left(  t\right)  \text{.} \label{alphae}%
\end{equation}
Noting that $\dot{X}\left(  t\right)  X^{\ast}\left(  t\right)  \overset
{\text{def}}{=}\operatorname{Re}\left[  \dot{X}\left(  t\right)  X^{\ast
}\left(  t\right)  \right]  +i\operatorname{Im}\left[  \dot{X}\left(
t\right)  X^{\ast}\left(  t\right)  \right]  $, integration of Eq.
(\ref{alphae}) yields%
\begin{equation}
\alpha\left(  t\right)  =\frac{\hslash}{\sqrt{\hslash^{2}+\left\vert X\left(
t\right)  \right\vert ^{2}}}\exp\left[  -\frac{i}{\hslash}\int_{0}^{t}%
\Omega\left(  t^{\prime}\right)  dt^{\prime}-i\int_{0}^{t}\frac
{\operatorname{Im}\left[  \dot{X}\left(  t^{\prime}\right)  X^{\ast}\left(
t^{\prime}\right)  \right]  }{\hslash^{2}+\left\vert X\left(  t^{\prime
}\right)  \right\vert ^{2}}dt^{\prime}\right]  \text{.} \label{alphaf}%
\end{equation}
In conclusion, once the auxiliary \emph{complex} function $X\left(  t\right)
$ and the \emph{real} longitudinal field $\Omega\left(  t\right)  $ have been
chosen, we can compute the probability amplitude $\alpha\left(  t\right)  $
from Eq. (\ref{alphaf}), the transversal field $\omega\left(  t\right)  $ from
Eq. (\ref{omegaf}), and finally, the probability amplitude $\beta\left(
t\right)  $ from\ Eq. (\ref{beta}). Clearly, once we have $\alpha\left(
t\right)  $ and $\beta\left(  t\right)  $, we can compute the transition
probability $\mathcal{P}_{\left\vert s\right\rangle \rightarrow\left\vert
w\right\rangle }\left(  t\right)  $ in Eq. (\ref{TP1}).

\end{document}